\documentclass[12pt,preprint]{aastex}
\def \agile {AGILE}
\def \egret {EGRET}

\def \igr {INTEGRAL}
\def \swi {{\it Swift}}
\def \rem {REM}
\def \webt {WEBT}
\def \cgro {CGRO}
\def \degmark{^\circ}
\def \ergsec{\hbox{erg s$^{-1}$}}

\def \phcmsec{\hbox{photons cm$^{-2}$ s$^{-1}$}}

\def \gray {$\gamma$-ray }
\def \source {\hbox{3C~454.3}}
\slugcomment{To appear in ApJ}

\shorttitle{Multiwavelength observations of the {\it Crazy Diamond} 3C~454.3}
\shortauthors{Vercellone et al.}

\begin{document}
\title{Multiwavelength observations of 3C~454.3. \\
  I. The AGILE 2007 November campaign on the ``{\it Crazy Diamond}''}
\author{S.~Vercellone\altaffilmark{1,*}, 
  A.W.~Chen\altaffilmark{1,2},
  V.~Vittorini\altaffilmark{3}, 
  A.~Giuliani\altaffilmark{1}, 
  F.~D'Ammando\altaffilmark{3,4},
  M.~Tavani\altaffilmark{3,4},
  I.~Donnarumma\altaffilmark{3}, 
  G.~Pucella\altaffilmark{3}, 
  C.M.~Raiteri\altaffilmark{5}, 
  M.~Villata\altaffilmark{5},
  W.P.~Chen\altaffilmark{6},
  G.~Tosti\altaffilmark{7},
  D.~Impiombato\altaffilmark{7},
  P.~Romano\altaffilmark{8},
  A.~Belfiore\altaffilmark{9},
  A.~De~Luca\altaffilmark{1,9,10},
  G.~Novara\altaffilmark{11},
  F.~Senziani\altaffilmark{9,1},
  A.~Bazzano\altaffilmark{3},
  M.T.~Fiocchi\altaffilmark{3},
  P.~Ubertini\altaffilmark{3},
  A.~Ferrari\altaffilmark{2,12},
  A.~Argan\altaffilmark{3}, 
  G.~Barbiellini\altaffilmark{6},
  F.~Boffelli\altaffilmark{10,11},
  A.~Bulgarelli\altaffilmark{13},
  P.~Caraveo\altaffilmark{1}, 
  P.W.~Cattaneo\altaffilmark{10}, 
  V.~Cocco\altaffilmark{3},
  E.~Costa\altaffilmark{3},
  E.~Del Monte\altaffilmark{3}, 
  G.~De Paris\altaffilmark{3},
  G.~Di Cocco\altaffilmark{13},
  Y.~Evangelista\altaffilmark{3},
  M.~Feroci\altaffilmark{3},
  M.~Fiorini\altaffilmark{1},
  F.~Fornari\altaffilmark{1},
  T.~Froysland\altaffilmark{3,14},
  F.~Fuschino\altaffilmark{13}, 
  M.~Galli\altaffilmark{15},
  F.~Gianotti\altaffilmark{13}, 
  C.~Labanti\altaffilmark{13}, 
  I.~Lapshov\altaffilmark{3}, 
  F.~Lazzarotto\altaffilmark{3}, 
  P.~Lipari\altaffilmark{16}, 
  F.~Longo\altaffilmark{17}, 
  M.~Marisaldi\altaffilmark{13},
  S.~Mereghetti\altaffilmark{1},
  A.~Morselli\altaffilmark{14}, 
  A.~Pellizzoni\altaffilmark{1}, 
  L.~Pacciani\altaffilmark{3},
  F.~Perotti\altaffilmark{1},
  P.~Picozza\altaffilmark{14}, 
  M.~Prest\altaffilmark{18},
  M.~Rapisarda\altaffilmark{19},
  A.~Rappoldi\altaffilmark{10},
  P.~Soffitta\altaffilmark{3}, 
  M.~Trifoglio\altaffilmark{13},
  A.~Trois\altaffilmark{3}, 
  E.~Vallazza\altaffilmark{6},
  A.~Zambra\altaffilmark{1},
  D.~Zanello\altaffilmark{16},
  C.~Pittori\altaffilmark{20}, 
  F.~Verrecchia\altaffilmark{20},
  P.~Santolamazza\altaffilmark{20},
  B.~Preger\altaffilmark{20},
  D.~Gasparrini\altaffilmark{20}, 
  S.~Cutini\altaffilmark{20},
  P.~Giommi\altaffilmark{20}, 
  S.~Colafrancesco\altaffilmark{20},
  L.~Salotti\altaffilmark{21}
}
\altaffiltext{1}{INAF/IASF--Milano, Via E.~Bassini 15, I-20133 Milano, Italy}
\altaffiltext{2}{CIFS--Torino, Viale Settimio Severo 3, I-10133, Torino, Italy}
\altaffiltext{3}{INAF/IASF--Roma, Via del Fosso del Cavaliere 100, 
  I-00133 Roma, Italy}
\altaffiltext{4}{Dip. di Fisica, Univ. ``Tor Vergata'', Via della Ricerca 
  Scientifica 1, I-00133 Roma, Italy}
\altaffiltext{5}{INAF/OATo, Via Osservatorio 20, I-10025 Pino Torinese, Italy}
\altaffiltext{6}{Institute of Astronomy, National Central University, Taiwan}
\altaffiltext{7}{Dip. di Fisica, Univ. di Perugia, Via Pascoli, I-06123 Perugia, Italy}
\altaffiltext{8}{INAF/IASF--Palermo, Via U. La Malfa 153, I-90146 Palermo, Italy}
\altaffiltext{9}{Ist. Univ. di Studi Superiori, V.\ le Lungo Ticino 56, I-27100 Pavia, Italy}
\altaffiltext{10}{INFN--Pavia, Via Bassi 6, I-27100 Pavia, Italy}
\altaffiltext{11}{Dip. di Fisica Nucleare e Teorica, Univ. degli Studi di Pavia, Via Bassi 6, I-27100 Pavia, Italy}
\altaffiltext{12}{Dip. di Fisica , Univ. di Torino, Via P. Giuria 1, I-10125 Torino, Italy}
\altaffiltext{13}{INAF/IASF--Bologna, Via Gobetti 101, I-40129 Bologna, Italy}
\altaffiltext{14}{INFN--Roma ``Tor Vergata'', Via della Ricerca Scientifica 1, I-00133 Roma, Italy}
\altaffiltext{15}{ENEA--Bologna, Via Martiri di Monte Sole 4, I-40129 Bologna, Italy}
\altaffiltext{16}{INFN--Roma ``La Sapienza'', Piazzale A. Moro 2, I-00185 Roma, Italy}
\altaffiltext{17}{Dip. di Fisica and INFN, Via Valerio 2, I-34127 Trieste, Italy}
\altaffiltext{18}{Dip. di Fisica, Univ. dell'Insubria, Via Valleggio 11, I-22100 Como, Italy}
\altaffiltext{19}{ENEA--Roma, Via E. Fermi 45, I-00044 Frascati (Roma), Italy}
\altaffiltext{20}{ASI--ASDC, Via G. Galilei, I-00044 Frascati (Roma), Italy}
\altaffiltext{21}{ASI, Viale Liegi 26 , I-00198 Roma, Italy}
\altaffiltext{*}{Email: \texttt{stefano@iasf-milano.inaf.it}}

 \begin{abstract}
   {We report on a multiwavelength observation of the blazar
   3C~454.3 (which we dubbed {\it crazy diamond}) carried out on
   November 2007 by means
   of the astrophysical satellites \agile{}, \igr{}, \swi{}, the
   \webt{} Consortium, and the optical-NIR telescope \rem{}.}
   Thanks to the
   wide field of view (FoV) of the \agile{} satellite and its 
   prompt alert dissemination to other observatories, we obtained 
   a long (three weeks), almost continuous \gray coverage of 
   the blazar 3C~454.3 across fourteen decades of energy.
   {This broad-band monitoring allows us to study in great detail
   light curves, correlations, time-lags and spectral energy 
   distributions (SEDs) during different physical states.}
   {Gamma-ray data were collected  during an \agile{}
   pointing towards the Cygnus Region. Target of
   Opportunity (ToO) observations were performed
   to follow-up the \gray observations in the soft and hard X-ray 
   energy bands. 
   Optical data were acquired continuously by means of a pre-planned 
   \webt{} campaign and through a \rem{} ToO repointing.}
   {\source{} is detected at a $\sim 19-\sigma$ level during the 3-week
   observing period, with an average flux above 100~MeV of 
   $ F_{\rm E>100MeV} = (170 \pm 13) \times 10^{-8}$\,\phcmsec. The
   \gray spectrum can be fit with a single power-law with photon
   index $\Gamma_{\rm GRID} = 1.73 \pm 0.16$ between 
   100~MeV and 1~GeV.
   We detect significant day-by-day variability of the \gray emission 
   during our observations, and
   we can exclude that the fluxes are constant at the 99.6\%
   ($\sim 2.9 \sigma$) level.
   The source was detected typically around 40 degrees off-axis, 
   and it was substantially off--axis in the field of view 
   of the AGILE hard X-ray imager.
   However, a 5-day long ToO observation by \igr{} detected \source{}
   at an average flux of about
   $F_{\rm 20-200\,keV} = 1.49 \times 10^{-3}$\,\phcmsec with an
   average photon index of $\Gamma_{\rm IBIS} = 1.75 \pm 0.24$ 
   between 20--200~keV.
   \swi{} also detected \source{} with a flux in the
   0.3--10~keV energy band in the range
   $(1.23-1.40) \times 10^{-2}$\,\phcmsec{} and a photon index
   in the range $\Gamma_{\rm XRT} = 1.56-1.73$.
   In the optical band, both WEBT and REM show an extremely
   variable behavior in the $R$ band.
   A correlation analysis based on the entire data set is consistent 
   with no time-lags between the \gray and the optical flux variations.}
   {Our simultaneous multifrequency observations strongly indicate 
   that the dominant emission
   mechanism between 30~MeV and 30~GeV is dominated by 
   inverse Compton scattering of relativistic electrons in the jet on the
   external photons from the broad line region.}
\end{abstract}

%
%
       %%%%%%%%%%%%%%%%%%%%%%%%%%%%%%%%%%%%%%%%%%%%%%%%%%%%%%%%%%%%%%%%%%%%
                       \section{Introduction} \label{3c454:intro}
       %%%%%%%%%%%%%%%%%%%%%%%%%%%%%%%%%%%%%%%%%%%%%%%%%%%%%%%%%%%%%%%%%%%%
%
%
Among active galactic nuclei (AGNs), blazars show intense and variable
\gray emission above 100~MeV \citep{Hartman1999:3eg}. Variability timescale
can be as short as few days, or last few weeks.
They emit across several decades of energy, from the radio to the TeV
energy band.
Blazar spectral energy distributions (SEDs) are typically 
double humped with a first peak occurring in
the IR/Optical band in the so-called {\it red blazars} (including
Flat Spectrum Radio Quasars, FSRQs, and Low-energy peaked BL Lacs, LBLs)
and at UV/X-rays in the so-called {\it blue blazars} (including
High-energy peaked BL Lacs, HBLs) and it is commonly 
interpreted as synchrotron radiation
from high-energy electrons in a relativistic jet. The second SED
component, which peaks at MeV--GeV energies in
red blazars and at TeV energies in blue blazars, is commonly
interpreted as Inverse Compton (IC) scattering of soft seed photons by
relativistic electrons.
A recent review of the blazar emission mechanisms and energetics
is given in \cite{Celotti2008:blazar:jet}.
Alternatively, the blazar SED can be explained in the framework of the
hadronic models, where the relativitic protons in the jet are the primary 
accelerated particles, emitting \gray radiation by means of photo-pair and 
photo-pion production (see \citealt{Mucke2001:hadronic,Mucke2003:hadronic} 
for a recent review on hadronic models).

Multiwavelength studies of variable \gray blazars have been
carried out since the beginning of the 1990s, thanks to the 
\cgro{} Observatory. 
Nevertheless, only a few objects 
were detected on a timescale of about two weeks in the
\gray energy band, and simultaneously monitored at different
energies, obtaining a multifrequency coverage.
Among the FSRQs detected at energies above 100~MeV by 
the \egret{} telescope on-board the Compton Gamma-Ray Observatory
\citep{Hartman1999:3eg}, 
\source{} (PKS~2251$+$158; $z=0.859$) is certainly one of the most active
at high energy. In the \egret{} era, it was detected in 1992
during an intense \gray flaring episode 
\citep{Hartman1992:3C454iauc, Hartman1993:3C454_EGRET}
when its flux $F_{\rm E>100MeV}$ was observed to vary within the range
$(0.4-1.4) \times 10^{-6}$\,photons\,cm$^{-2}$\,s$^{-1}$. In 1995, a 2-week 
campaign detected a \gray flux $< 1/5$ of its historical maximum
\citep{Aller1997:3C454_EGRET}.

In 2005, \source{} underwent a  major flaring activity in almost 
all energy bands (see \citealt{Giommi2006:3C454_Swift}).
In the optical, it reached $R=12.0$\,mag \citep{vil06} 
and it was detected by \igr{} at a flux\footnote{Assuming
a Crab-like spectrum.} level of
$\sim 3 \times 10^{-2}$\,photons\,cm$^{-2}$\,s$^{-1}$ in the 3--200~keV 
energy band \citep{Pian2006:3C454_Integral}. 
Since the detection of the exceptional 2005 outburst, several monitoring 
campaigns were carried out to follow the source multifrequency behavior 
\citep{vil06,vil07,rai07,rai08a,rai08b}. During the last of these campaigns, 
3C~454.3 underwent a new 
optical brightening in mid July 2007, which triggered observations at all 
frequencies.

In July 2007, \citet[][hereafter V08]{Vercellone2008:3C454_ApJ} reported the
highest \gray flare from \source{}. During the period 2007 July 24--30 
the average \gray flux was
$F_{\rm E>100MeV} = (280 \pm 40) \times 10^{-8}$\,\phcmsec, 
about a factor of two higher 
than in 1995. No emission was detected by Super-AGILE in the energy 
range 20--60~keV, with a 3-$\sigma$ upper limit of
$2.3 \times 10^{-3}$\,\phcmsec.

By means of \agile{} preliminary flux estimate \citep{Vercellone2007:atel1160},
\cite{Ghisellini2007:3C454:SED} compared the \source{} SEDs
obtained during three multiwavelength campaigns 
(2000, 2005 and 2007). The 2007 data show that the \gray high state
occurred during a weaker optical/X-ray flux compared with the 2005
flare.

In this paper (Paper I) we present the results of a multiwavelength campaign 
on \source{} during a long-lasting \gray activity period between
2007 November 10 and  December 1. 
Preliminary \gray results were distributed in 
\citet{Chen2007:ATel3C454}, while radio-to-optical and UV data
were published in \citet{rai08a}. A companion paper (Paper II, 
Donnarumma et al., in preparation) will describe the \agile{} multiwavelength
campaign during December 2007.
In Sect.~\ref{3c454:ml} we present the multiwavelength campaign
on \source{}; in Sect.~\ref{3c454:grid} through~\ref{3c454:optical} 
we present the \agile/GRID,
\igr/IBIS, \swi/XRT, \webt{} and \rem{} data analysis, respectively;
in Sec.~\ref{3c454:sim:data} we present the simultaneous
multiwavelength light-curves and SEDs.
In Sect.~\ref{3c454:disc} and~\ref{3c454:conc} we discuss
our results and draw our conclusions.
Throughout this paper the quoted uncertainties are given at the 
1--$\sigma$ level, unless otherwise stated.
%
%
       %%%%%%%%%%%%%%%%%%%%%%%%%%%%%%%%%%%%%%%%%%%%%%%%%%%%%%%%%%%%%%%%%%%%
                  \section{The Multiwavelength Campaign} \label{3c454:ml}
       %%%%%%%%%%%%%%%%%%%%%%%%%%%%%%%%%%%%%%%%%%%%%%%%%%%%%%%%%%%%%%%%%%%%
%
%
In 2007 November AGILE began pointing \source{} at high off-axis angle
(about $40\degmark$). Nevertheless, in a few days \source{} was 
detected at more than 5-$\sigma$ \citep{Chen2007:ATel3C454},
exhibiting variable activity on a day time-scale \citep{Pucella2007:ATel3C454}.
Immediately after the source detection, a multiwavelength campaign
started. 
\agile{} data were collected during two different periods, the
first ranging between  2007-11-10 12:17 UT and 2007-11-25 10:57 UT and
the second between 2007-11-28 12:05 UT and 2007-12-01 11:39 UT,
for a total of about 592\,ks.
The three-day gap between them was due to a pre-planned GRID 
calibration activity. 
\igr{} data were collected during a dedicated ToO on revolutions 
623 (between 2007-11-20 03:35 UT and 2007-11-22 08:46 UT) and 
624 (between 2007-11-22 20:45 UT and
2007-11-24 15:50 UT), for a total of about 300\,ks, while
\swi{}/XRT data were obtained during several ToO pointings for a total 
of about 10\,ks.
\webt{} data (radio to optical) as well as \swi/UVOT data were published 
in \cite{rai08a}, while
\rem{} data were acquired following a ToO request.
In both cases, optical data were acquired continuously during the whole
\agile{} campaign.

%
%
       %%%%%%%%%%%%%%%%%%%%%%%%%%%%%%%%%%%%%%%%%%%%%%%%%%%%%%%%%%%%%%%%%%%%
                       \section{\agile{} data} \label{3c454:grid}
       %%%%%%%%%%%%%%%%%%%%%%%%%%%%%%%%%%%%%%%%%%%%%%%%%%%%%%%%%%%%%%%%%%%%
%
%
%%%%%--------------------------------------------
     \subsection{Data Reduction and Analysis} \label{3c454:grid:analysis}
%%%%%--------------------------------------------
%
The AGILE satellite \citep{Tavani2008_agile_nima,Tavani2008:Missione},
a mission of the Italian Space Agency (ASI) devoted to high-energy
astrophysics, is currently the only space mission capable of 
observing cosmic sources simultaneously in the energy bands 
18--60~keV and 30~MeV--50~GeV. The satellite
was launched on 2007 April 23 by the Indian
PSLV-C8 rocket from the Satish Dhawan Space Center SHAR,
Sriharikota. 

The AGILE scientific Instrument is very compact and combines four
active detectors yielding broad-band coverage from hard X-rays
to gamma-rays: 
a Silicon Tracker~\citep[ST;][30~MeV--50~GeV]{Prest2003:agile_st},
a co-aligned coded-mask hard X-ray imager
\citep[SA;][18--60~keV]{Feroci2007:agile_sa}, a non-imaging CsI
Mini--Calorimeter~\citep[MCAL;][0.3--100~MeV]{Labanti2006:agile_mcal},
and a segmented Anti-Coincidence System~\citep[ACS;][]{Perotti2006:agile_ac}.
Gamma-ray detection is obtained by the combination of ST, 
MCAL and ACS; these three detectors form the AGILE Gamma-Ray Imaging 
Detector (GRID). 

Level--1 AGILE-GRID data were analyzed using the AGILE Standard Analysis 
Pipeline (see~V08 for a detailed discussion of
the AGILE data reduction). Since \source{} was at high off-axis angle, an ad-hoc
\gray analysis was performed. We used \gray events filtered by means
of the \texttt{FT3ab$\_$2} \agile{} Filter pipeline. 
Counts, exposure, and Galactic background \gray maps are created with
a bin-size of $0.\!\!^{\circ}25 \times 0.\!\!^{\circ}25$\,,
for $E \ge 100$\,MeV. Since the source was at $40^{\circ}$ off-axis,
all the maps were generated including all events collected up to
$60^{\circ}$ off-axis.
We rejected all the \gray events whose reconstructed
directions form angles with the satellite-Earth vector smaller
than $80^{\circ}$ (\texttt{albrad=80}), 
reducing the \gray Earth albedo contamination
by excluding regions within $\sim 10^{\circ}$ from the
Earth limb.
The most recent versions (\texttt{BUILD-15}) of the Calibration files,
which will be publicly available at the ASI Science Data Centre
(ASDC\footnote{\texttt{http://agile.asdc.asi.it}}) site,
and of the \gray diffuse emission model \citep{Giuliani2004:diff_model}
were used. 
The first step consists in running the AGILE Source Location task in order 
to derive the most plausible location of the source. In a second step,
we ran the AGILE Maximum Likelihood Analysis (\texttt{ALIKE}) using
a radius of analysis of 10$^{\circ}$\,,
and the best guess position derived
in the first step. The particular choice of the radius of analysis parameter
is dictated to avoid any possible contamination by very off-axis residual
particle events.
%
%
%%%%%--------------------------------------------
     \subsection{Results} \label{3c454:grid:results}
\begin{figure}[ht]
\resizebox{\hsize}{!}{\includegraphics[angle=0]{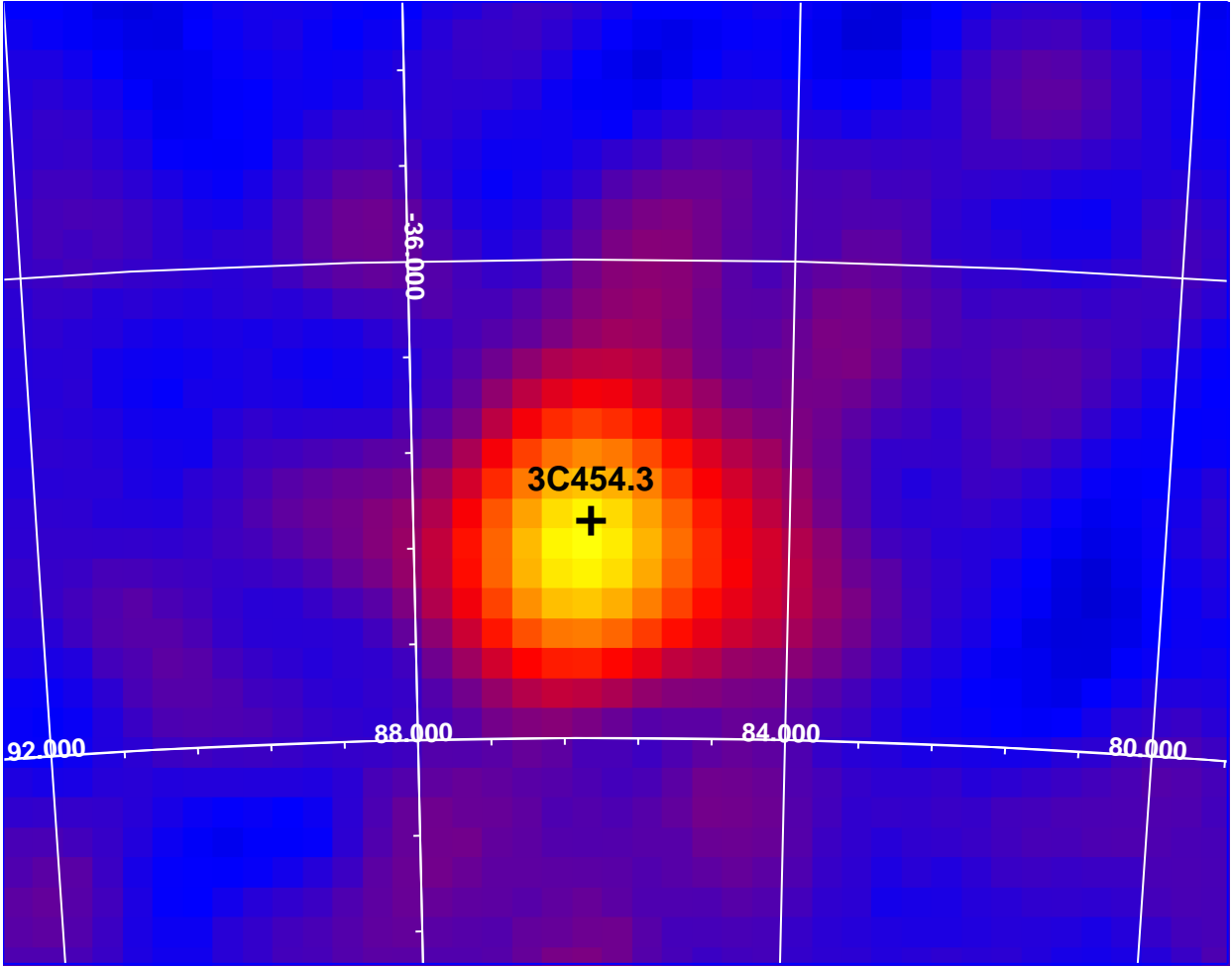}}
  \caption{Gaussian-smoothed intensity map 
    ($\sim 10^{\circ} \times 08^{\circ}$)
    in Galactic coordinates integrated over the whole 
    observing period (2007 November 10 12:17 UT -- 2007 December 01 11:39 UT). 
    The cross symbol is located at the \source{}
    radio coordinates.
    [{\it See the electronic edition of the Journal for a color version of 
      this figure.}]
  }
  \label{3c454:fig:map}
\end{figure}

\begin{figure}[ht]
\resizebox{\hsize}{!}{\includegraphics[angle=0]{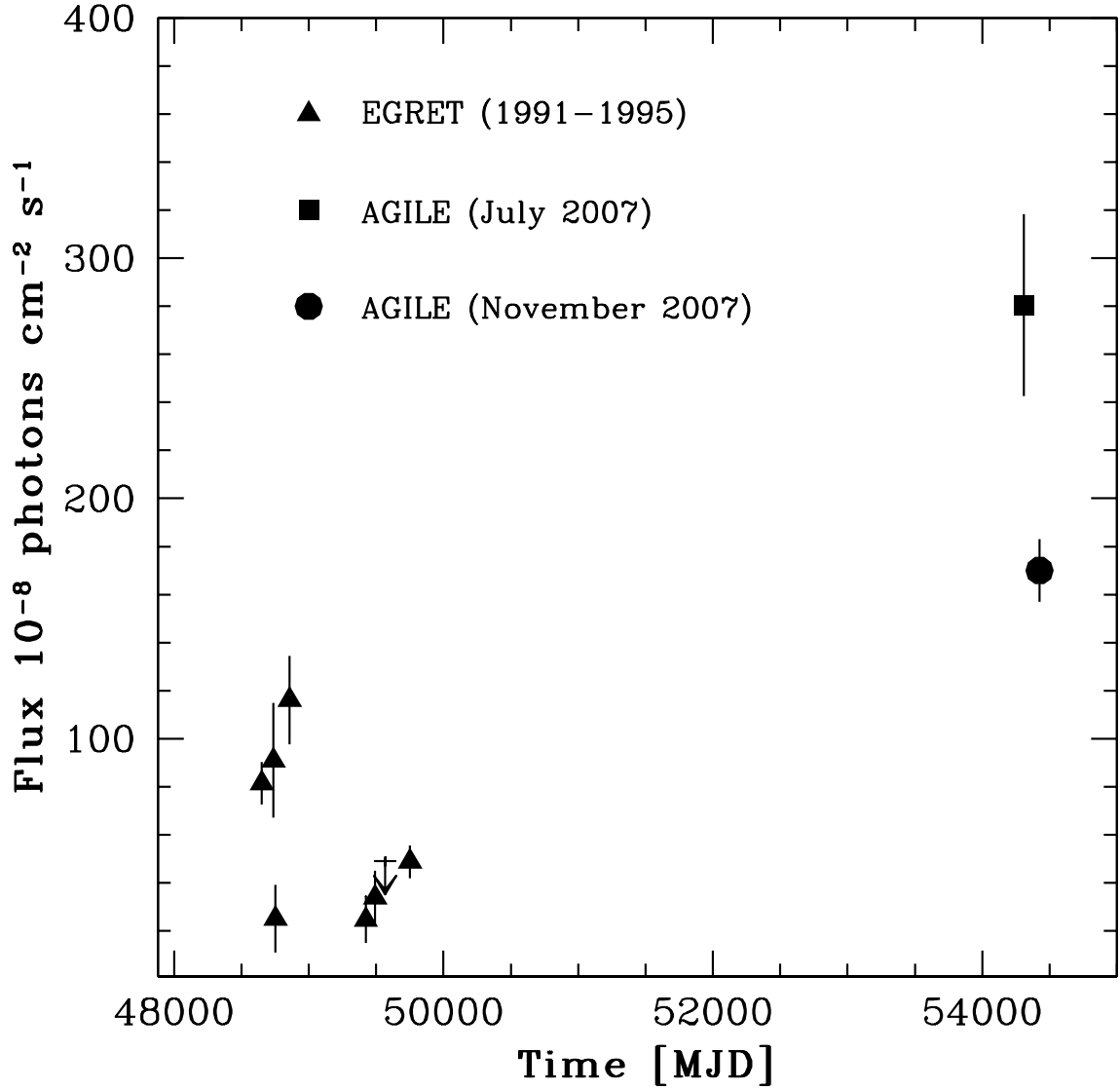}}
  \caption{ EGRET (triangles) and 
    AGILE--GRID (square and circle) gamma-ray
    light curve in units of $10^{-8}$\,\phcmsec. 
    EGRET data are from \citet{Hartman1999:3eg}.
    AGILE July 2007 data are from V08.
  }
  \label{3c454:fig:egret_agile}
\end{figure}

%%%%%--------------------------------------------
%
Figure~\ref{3c454:fig:map} shows a Gaussian-smoothed intensity map 
($\sim 10^{\circ} \times 08^{\circ}$) in Galactic coordinates
integrated over the whole observing period, using the selections described 
in~\S\ref{3c454:grid:analysis}.
The source detection significance is 
19-$\sigma$ and the average \gray flux above 100 MeV for 
the whole period is
$ F_{\rm E\,>\,100\,MeV} = (170 \pm 13) \times 10^{-8}$\,\phcmsec,
as derived from the \agile{} Maximum Likelihood Code analysis.
We note that the average \gray flux computed over the three week
campaign is lower than 
$F^{\rm July}_{\rm E>100\,MeV}=(280 \pm 40) \times 10^{-8}$\,\phcmsec\,,
observed during the flaring episode in July 2007, and computed
during only a six day observation. Nevertheless, 
Figure~\ref{3c454:fig:egret_agile} shows that the
current average flux is still higher than those observed during the
\egret{} era.
The smaller errors on the \agile{} November data with respect to
the July data are due to both the higher statistics (323 versus 
101 counts collected in November and in July, respectively), 
and to the more accurate calibration response files.

\begin{figure}[ht]
\resizebox{\hsize}{!}{\includegraphics[angle=-90]{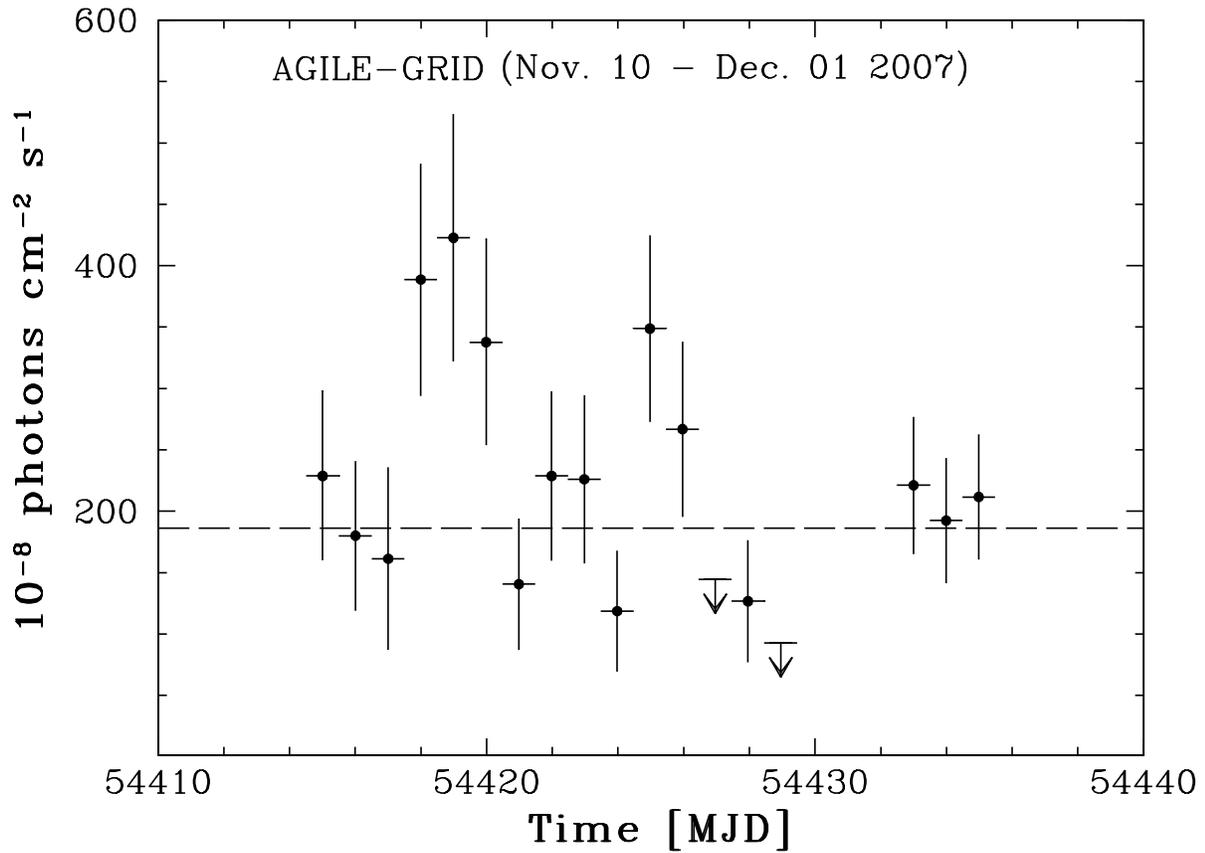}}
  \caption{AGILE--GRID \gray light-curve at 
       $\approx 1$-day resolution for E$>$100~MeV in units of 
      $10^{-8}$\,\phcmsec. The downward arrows represent 2-$\sigma$
      upper-limits. The dashed line represents the weighted mean flux.
  }
  \label{3c454:fig:gammalc}
\end{figure}

Figure~\ref{3c454:fig:gammalc} shows the \gray light-curve
at 1-day resolution for photons above 100~MeV. 
We note that \source{} is detected at a 3-$\sigma$ level
during almost the whole period on a 1-day timescale; this clearly
indicates strong \gray flaring activity. 

The average \gray flux as well as the daily values
of the 18 days were derived according
to the \gray analysis procedure described in 
\citet{Mattox1993:1633}.
First, the entire period was analyzed to determine the diffuse
gas parameters and then the source flux density was estimated
independently for each of the eighteen 1-day periods with the diffuse
parameters fixed at the values obtained in the first step.
Fitting the GRID fluxes to a constant model (the weighted
mean of the 1-day average flux values) yields
$F_{\rm wtd} = (186.3 \pm 14.6) \times 10^{-8}$\,\phcmsec.
Following \cite{McLaughlin1996:v:index}, we computed the $V$ variability
coefficient. The 2-$\sigma$ upper limits (UL) were properly treated,
assigning a value and a sigma equal to $UL/2$ \citep{Torres2001:variab:index}.
We obtain a $\chi^2=36.7$ for 17 degrees of freedom (d.o.f.);
therefore we can exclude that the fluxes are constant at the 99.6\%
($\sim 2.9 \sigma$) level, and we obtain a value for the variability
coefficient $V$ of 2.43.
A value of $V>1$ indicates that the source is variable within the 
observed period.

\begin{figure}[ht]
\resizebox{\hsize}{!}{\includegraphics[angle=0]{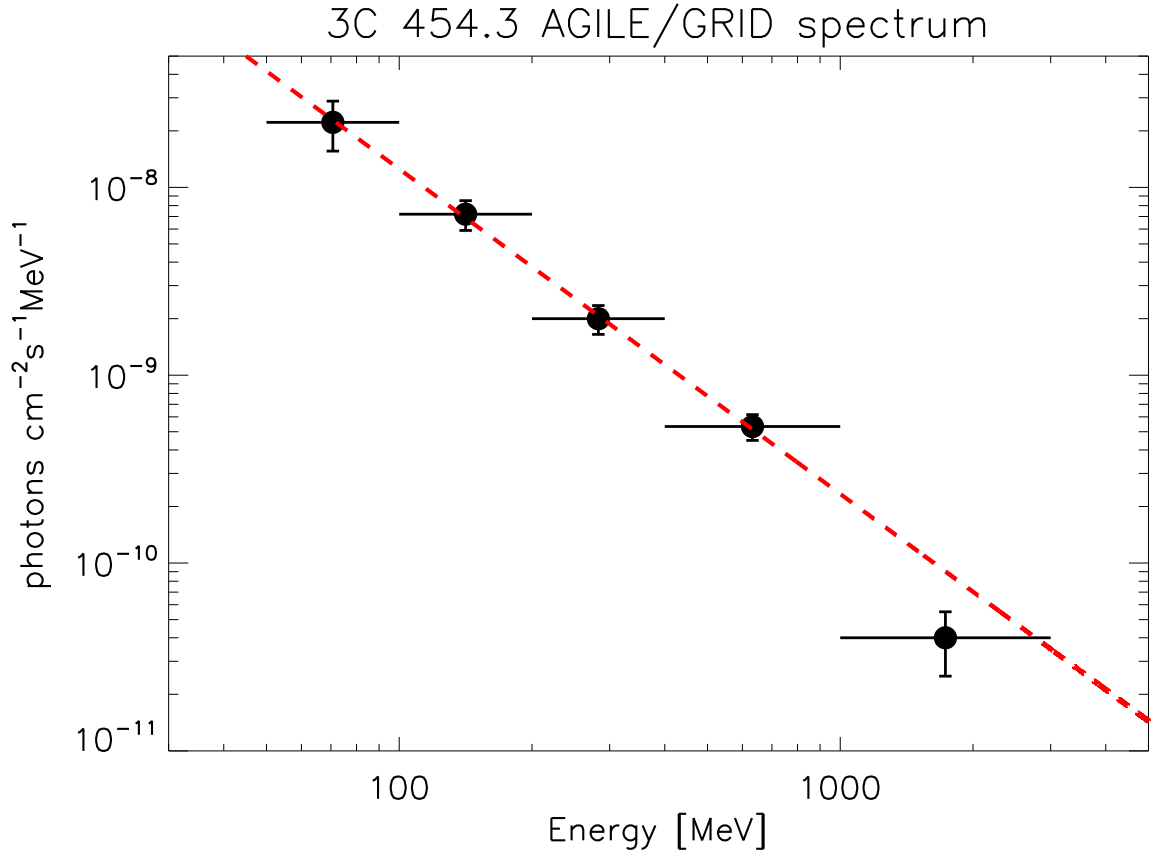}}
  \caption{AGILE--GRID average \gray spectrum. Three energy bins were
    considered: $100 < {\rm E} < 200$~MeV, $200 < {\rm E} < 400$~MeV,
    $400 < {\rm E} < 1000$~MeV.
    The red dashed line represents the best--fit power law model.
    [{\it See the electronic edition of the Journal for a color version of 
      this figure.}]
  }
  \label{3c454:fig:gammaspec}
\end{figure}

Figure~\ref{3c454:fig:gammaspec} shows the average \gray spectrum
derived over the entire observing period.
The average spectrum was obtained by computing the \gray flux in 
five energy bins over the entire observing period:
$50 < {\rm E} < 100$~MeV,
$100 < {\rm E} < 200$~MeV, $200 < {\rm E} < 400$~MeV, 
$400 < {\rm E} < 1000$~MeV, and $1000 < {\rm E} < 3000$~MeV. 
We fit the data by means of a simple power-law model and restricted
our fit to the most reliable energy range (100~MeV--1~GeV):
\begin{equation}\label{3c454:equ:defflux}
  F(E) = 
  3.61\times 10^{-5}
  \left( \frac{{\rm E}}{1\, {\rm MeV}}\right)^{-(1.73 \pm 0.16)}
  {\rm ph\,cm^{-2}\,s^{-1}\,MeV^{-1}}.
\end{equation}

Unfortunately, the source was located substantially
off-axis in the Super--AGILE FoV during the whole observation period,
resulting in a not particularly constraining upper
limit flux, being as high as $1.13\times 10^{-2}$\,\phcmsec{}
(50~mCrab).
%
       %%%%%%%%%%%%%%%%%%%%%%%%%%%%%%%%%%%%%%%%%%%%%%%%%%%%%%%%%%%%%%%%%%%%
                       \section{\igr{} data} \label{3c454:igr}
       %%%%%%%%%%%%%%%%%%%%%%%%%%%%%%%%%%%%%%%%%%%%%%%%%%%%%%%%%%%%%%%%%%%%
%
%
%%%%%--------------------------------------------
     \subsection{Data Reduction and Analysis} \label{3c454:igr:analysis}
%%%%%--------------------------------------------
%
The ESA \igr{} \gray{} Observatory, launched in 2002 October,
carries three co-aligned coded mask telescopes.
For the purpose of this paper we refer to 
data from the IBIS instrument \citep{Ubertini2003:ibis},
sensitive in the energy range
15\,keV--10\,MeV and with a FoV of 29$^{\circ}\times29^{\circ}$\,,
and in particular to the ISGRI lower energy detector layer.

All the observations are organized into un-interrupted 2000--3600 s 
long science windows (SCW): light curves and spectra were extracted 
for each individual SCW.
Wide-band spectra (from 17 to 150 keV) of the source were obtained using
data from IBIS instrument.
All the data were processed using the Off-line Scientific Analysis ({\texttt
OSA}) version 7.0 software released by the  \igr{} Scientific Data Centre.
\igr{} data were analyzed using FTOOLS and XSPEC11.3.2 in the {\tt Heasoft} 
package (v.6.4). We assumed a single power law model to fit the IBIS
data. 

\begin{figure}[ht]
\resizebox{\hsize}{!}{\includegraphics[angle=-90]{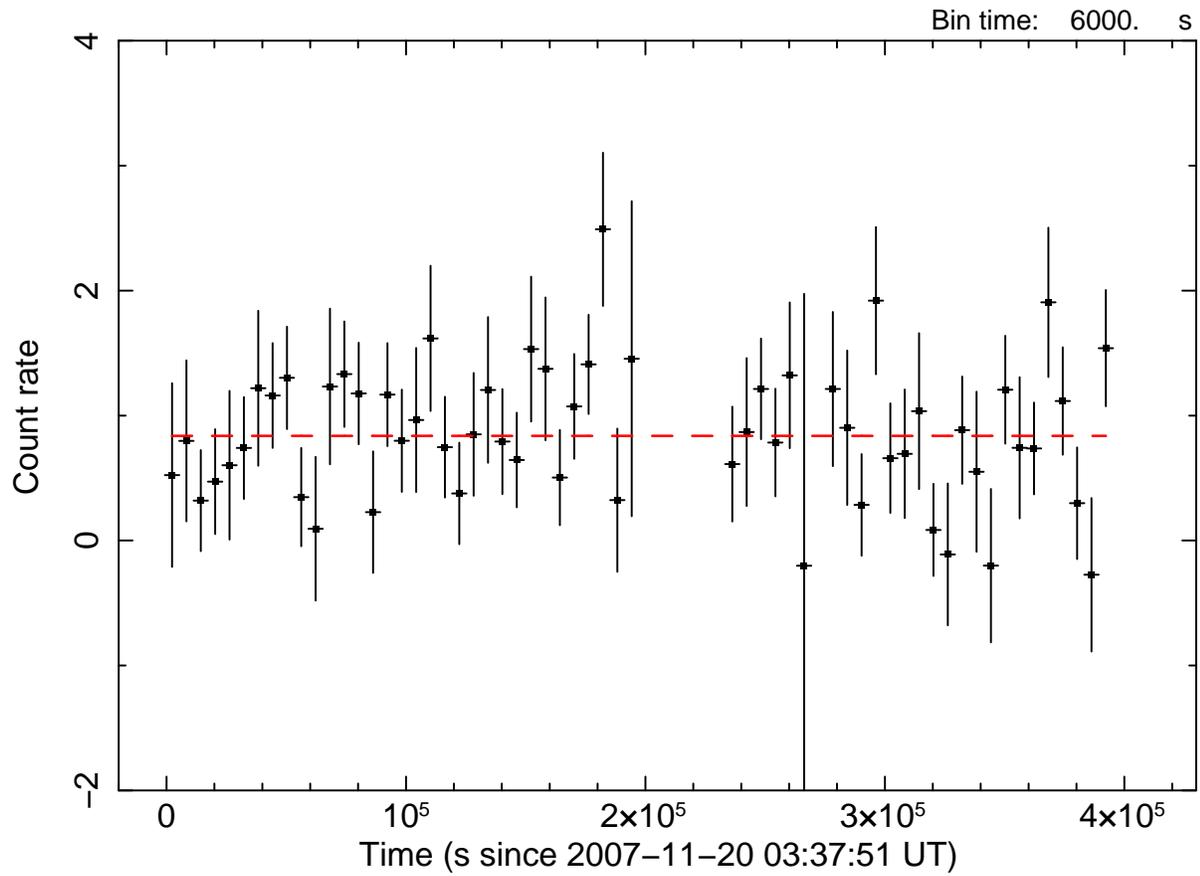}}
  \caption{INTEGRAL/IBIS light curve
    in the energy range 20--50~keV accumulated during the whole observation.
    The dashed line represents a fit with a constant model.
    [{\it See the electronic edition of the Journal for a color version of 
      this figure.}]
  }
  \label{3c454:fig:igr:lc2050}
\end{figure}

\begin{figure}[ht]
\resizebox{\hsize}{!}{\includegraphics[angle=-90]{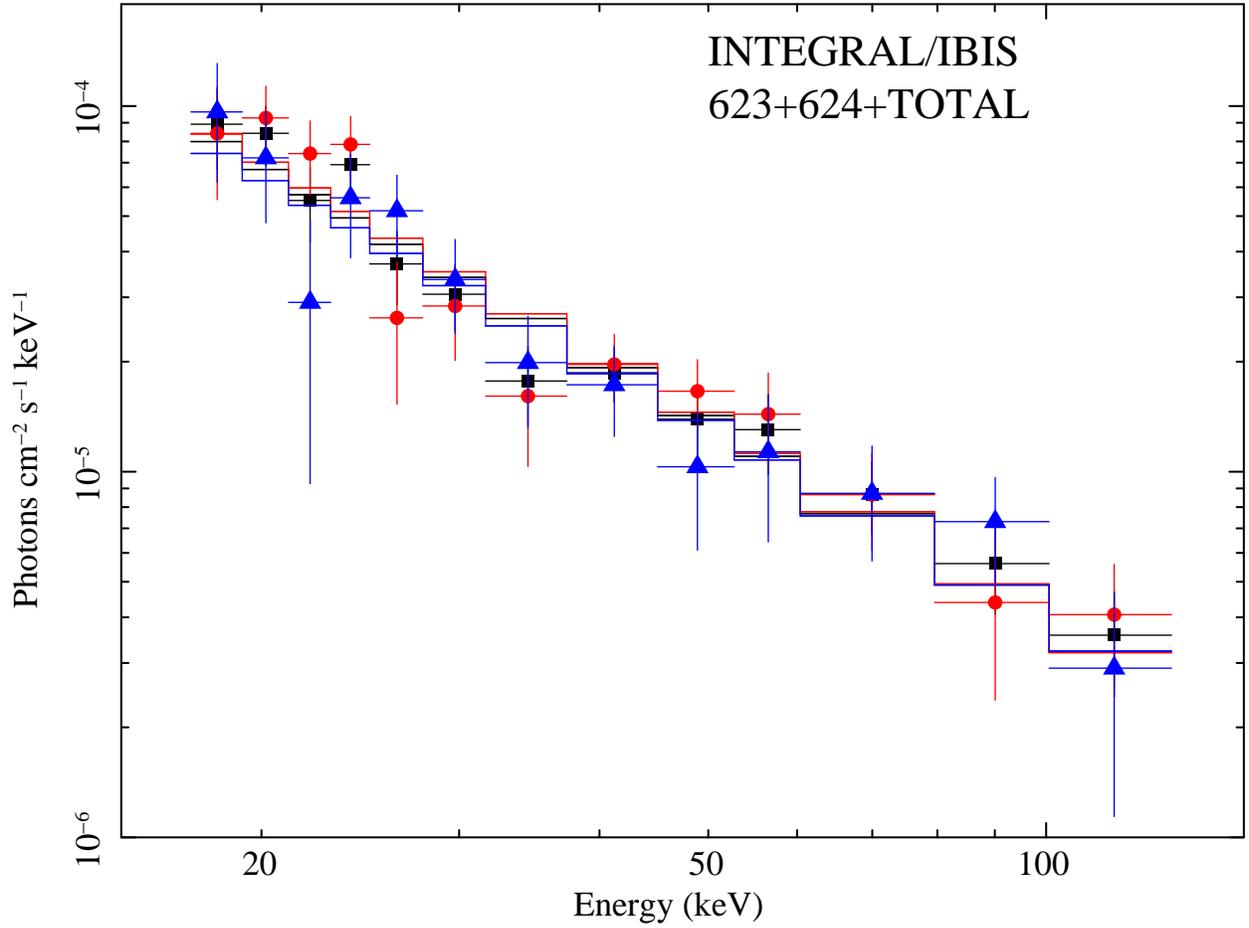}}
  \caption{INTEGRAL/IBIS spectra for revolution 623 (red circles),
    revolution 624 (blue triangles) and for the whole observation
    (black squares). 
    [{\it See the electronic edition of the Journal for a color version of 
      this figure.}]
  }
  \label{3c454:fig:igr:spectra_mt}
\end{figure}

Figure~\ref{3c454:fig:igr:lc2050} shows the \igr/IBIS light curve
in the energy range 20--50~keV accumulated during the whole observation.
The source does not show statistically significant flux variations.
Figure~\ref{3c454:fig:igr:spectra_mt} shows the INTEGRAL/IBIS spectra 
for revolution 623 (red circles),
revolution 624 (blue triangles) and for the whole observation
(black squares). 

Table~\ref{3c454:tab:igr:specfits} summarizes the \igr{}/IBIS 
spectral fit results.

%
%
       %%%%%%%%%%%%%%%%%%%%%%%%%%%%%%%%%%%%%%%%%%%%%%%%%%%%%%%%%%%%%%%%%%%%
                       \section{\swi{} data} \label{3c454:swift}
       %%%%%%%%%%%%%%%%%%%%%%%%%%%%%%%%%%%%%%%%%%%%%%%%%%%%%%%%%%%%%%%%%%%%
%
%
%%%%%--------------------------------------------
     \subsection{Data Reduction and Analysis} \label{3c454:swift:analysis}
%%%%%--------------------------------------------
The NASA \swi{} \gray Burst Mission \citep{Gehrels2004:swift}, launched 
in November 2004, has three co-aligned instruments: a coded-mask 
Burst Alert Telescope~\citep[BAT;][15--150\,keV]{Barthelmy2005SSRv:bat},
an X-ray Telescope~\citep[XRT;][0.2--10\,keV]{Burrows2005:grades}, and
an Ultraviolet/Optical Telescope~\citep[UVOT;][170--600\,nm]{Roming2005SSRv:UVOT}.

\swi{} data (Obs. ID 00031018) were collected by activating 
a \swi{} Cycle-3 Proposal (PI A.\ W.\ Chen) 
and by means of a dedicated ToO triggered by \agile{} (PI S. Vercellone).
The XRT data were processed with standard procedures 
({\tt xrtpipeline} v0.11.6),
adopting the standard filtering and screening criteria, and using FTOOLS 
in the {\tt Heasoft} package (v.6.4). 
The source count rate was low during the whole campaign,
thus we only considered photon counting data 
(PC) and further selected XRT event grades 0--12 \citep{Burrows2005:grades}. 
\swi{}/XRT data show an average count rate of $>0.5$ counts s$^{-1}$ and 
therefore pile-up correction was required. 
We extracted the source events from  an annular source extraction 
region with an inner radius of 2--3 pixels (estimated case by case 
by means of  the PSF fitting technique) and an outer radius of 30 pixels 
(1 pixel $\sim2\farcs37$).
To account for the background, we also extracted events within a
circular region centered on a region free from background sources 
and with radius of 80 pixels.
Ancillary response files were generated with {\tt xrtmkarf}, 
and account for different extraction regions, vignetting and 
PSF corrections. We used the spectral redistribution matrices 
v010 in the Calibration Database maintained by HEASARC.
\swi{}/XRT uncertainties are given at 
90\% confidence level 
for one interesting parameter (i.e., $\Delta \chi^2 =2.71$) 
unless otherwise stated.
 
\begin{figure}[ht]
\resizebox{\hsize}{!}{\includegraphics[angle=0]{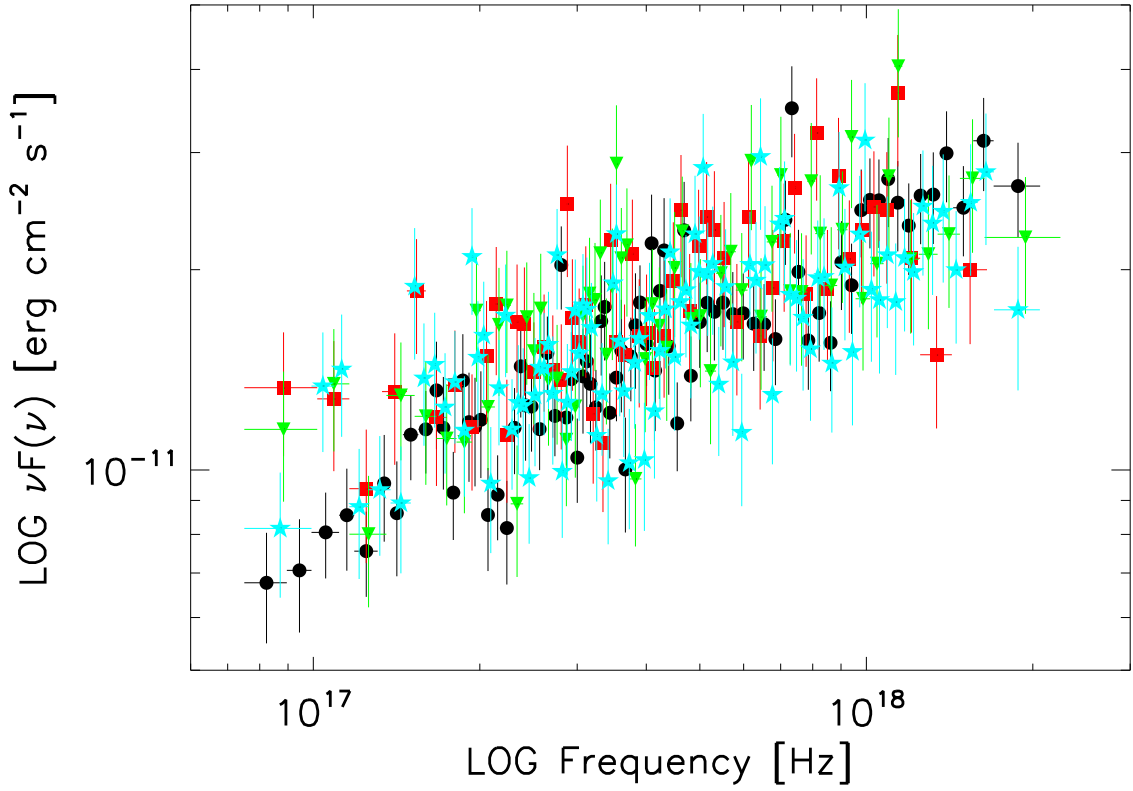}}
  \caption{\swi/XRT  0.3--10 keV spectra for segment 001 to 006:
    black circles (segm.~001), red squares (segm.~003+004), 
    green upside-down triangles (segm.~005), and cyan 
    stars (segm.~006).
    [{\it See the electronic edition of the Journal for a color version of 
      this figure.}]
  }
  \label{3c454:fig:xrt:spectra}
\end{figure}

Figure~\ref{3c454:fig:xrt:spectra} shows the 0.3--10 keV spectra 
for segment 001 to 006, where we summed segments 003 and 004
in order to have similar statistics as the others:
black circles (segm.~001), red squares (segm.~003+004), 
green upside-down triangles (segm.~005), and cyan 
stars (segm.~006).
Segment 002 was not considered since only 1~s of data were
recorded.
We first fit \swi{}/XRT spectra with an absorbed power law model,
named model A (\texttt{wabs*zwabs(powerlaw)} in \texttt{XSPEC 11.3.2}). 
Data were rebinned in order to have at least 20 counts per energy 
bin. The Galactic absorption was fixed to the value of
$N_{\rm H}^{\rm Gal} = 0.724 \times 10^{21}$\,cm$^{-2}$
\citep{Kalberla2005:nh}.
We considered, in addition to the Galactic absorption coefficient,
an extra absorption component, $N_{\rm H}^{\rm z}$, following
the results showed in \cite{Ghisellini2007:3C454:SED} and
\cite{rai07}.
A second spectral fit (model B) was performed considering
a simple power law with the absorption component as a free parameter
(\texttt{wabs*(powerlaw)}).
Table~\ref{3c454:tab:xrt:specfits} 
summarizes the most relevant spectral fit parameters.

We note that the $N_{\rm H}^{\rm z}$ component in model A
is not well constrained, while a simpler fit (model B)
characterized by a single power law model with free
absorption coefficient yields $N_{\rm H}$ values which are 
consistent within the uncertainties among different observations.
We also check the possible presence
of a double power law as reported in
\citet{rai08b} as follows.
We fixed the $N_{\rm H}$ to the value derived by
\citet{vil06}, $(1.34\pm0.05)\times 10^{21}$\,cm$^{-2}$ 
(based on {\it Chandra} data),
and fixed the hard power law index to the one we obtained 
by fitting the XRT data above 2\,keV. In most cases the harder 
power law component is not required,
as its normalization is consistent with zero.

\begin{figure}[ht]
\resizebox{\hsize}{!}{\includegraphics[angle=0]{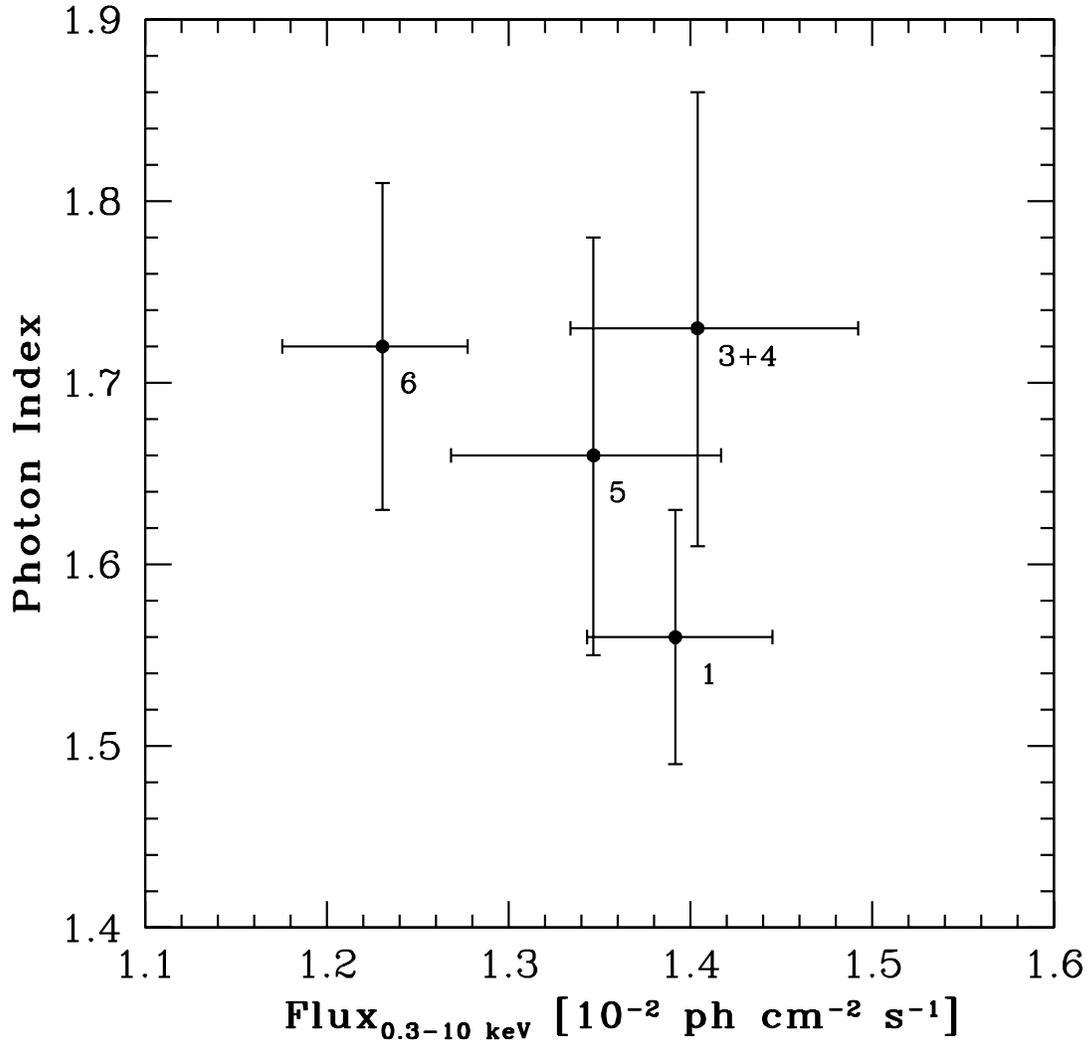}}
  \caption{\swi/XRT photon index versus the 0.3--10~keV flux.
    Numbers beneath each point represent the observing segment.
    Segment number 2 is missing since only 1 sec of data were
    recorded.
  }
  \label{3c454:fig:xrt:idxflux}
\end{figure}

Figure~\ref{3c454:fig:xrt:idxflux} shows the \swi/XRT photon index 
versus the 0.3--10~keV flux.
Numbers beneath each point represent the observing segment.

We analysed \swi{}/BAT Survey data in order to study the hard X-ray 
emission of \source{} and to investigate its evolution as a function of time.
We selected two time windows, the first between 2005 April 01 
and 2005 September 30 (when intense activity was recorded from
the target, see e.g. \citealt{Giommi2006:3C454_Swift}), 
the second between 2007 June 01 and 2007 December 31.
We considered all BAT observations with \source{} in the 
FoV. After a careful data selection, based on background rate,
pointing stability and several other data quality criteria 
(see \citealt{Senziani2007:bat:survey}), we ended up with 4824 observations
for a net exposure time of $\sim792$ ks in 6 months in 2005 and $\sim624$ ks 
in 7 months in 2007.

Here we provide a brief overview of the BAT Survey data analysis
procedure, while a detailed description will be addressed in a 
forthcoming paper (Belfiore et al., in preparation). 
Starting from Detector Plane Histogram (DPH) files, Detector Plane 
Images (DPI) in the 20--60 keV and 60--100 keV energy ranges were 
generated and were cleaned from hot pixels and noisy detectors.
Then, with the \texttt{HEASOFT} task {\em batfftimage} 
each DPI was deconvolved 
to obtain a background- and source-subtracted sky image of the BAT FoV. 
For each sky image an appropriate effective exposure map (weighted on the 
coded fraction) was generated, accounting for possible Earth/Moon occultations.
Then, such sky images were re-projected and stacked  
(weighting on effective exposure) to obtain monthly count rate 
maps, considering a small portion of the field around the target 
(a $3^{\circ}\times3^{\circ}$ region in local tangential
projection coordinates, TAN).

The count rates of \source{} were normalized to the Crab 
count rates and then converted to flux (ph cm$^{-2}$ s$^{-1}$)
assuming for the Crab the canonical power law spectrum (photon 
index $\Gamma=2.15$ and normalization of 10.4 ph cm$^{-2}$ s$^{-1}$
keV$^{-1}$ at 1 keV) and for \source{} a power law spectrum with
$\Gamma=1.7$, averaging the instrument response over the field
of view.
The November 2007 net exposure time is $\sim106$~ks for a flux
in the 20-60~keV energy band of $(1.07 \pm 0.19)\times 10^{-3}$\,
ph cm$^{-2}$ s$^{-1}$.

\begin{figure}[ht]
\resizebox{\hsize}{!}{\includegraphics[angle=0]{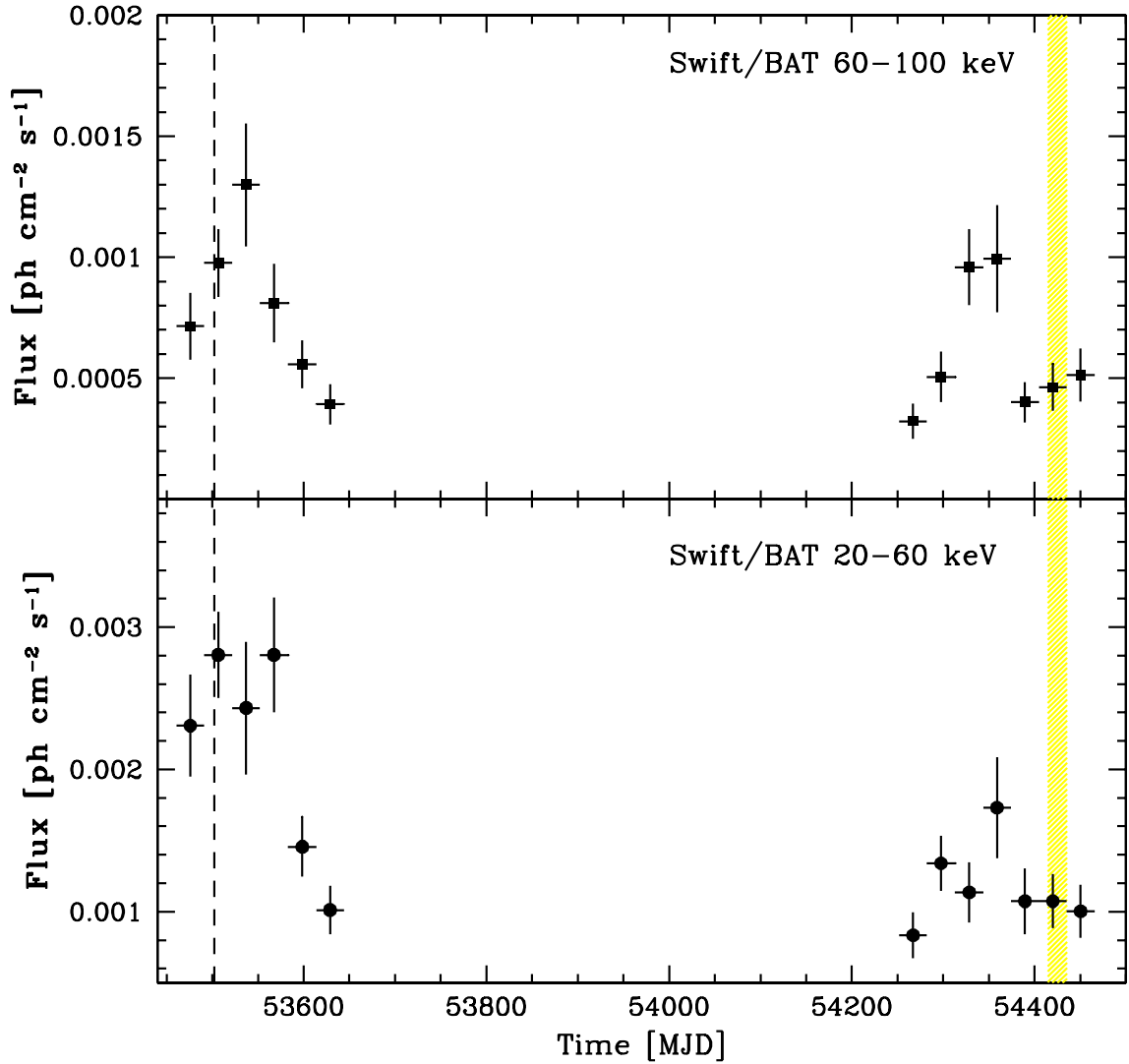}}
  \caption{Long-term {\it Swift}/BAT
    light curves in the 20--60~keV (bottom panel) and 60--100~keV
    (upper panel) energy range. The yellow vertical area marks the
    AGILE November campaign. The short-dashed line marks the epoch
    of the giant optical flare in 2005.
    [{\it See the electronic edition of the Journal for a color version of 
      this figure.}]
  }
  \label{3c454:fig:bat:lcurve}
\end{figure}

Figure~\ref{3c454:fig:bat:lcurve} shows the long-term {\it Swift}/BAT
light curves in the 20--60~keV (bottom panel) and 60--100~keV
(upper panel) energy range. The yellow vertical area marks the
AGILE November campaign. The {\it Swift}/BAT flux is in good
agreement with the flux derived from the whole \igr{}/IBIS campaign
in the same energy range, $F^{\rm IBIS}_{\rm 20-60 keV}=1.02\times 10^{-3}$\,
ph\,\,cm$^{-2}$ s$^{-1}$. The short-dashed line marks the epoch
of the giant optical flare in 2005 
\citep{Fuhrmann2006:3c454:opt,vil06}, 
when the hard X-ray flux was about twice higher than in November 2007.

%
       %%%%%%%%%%%%%%%%%%%%%%%%%%%%%%%%%%%%%%%%%%%%%%%%%%%%%%%%%%%%%%%%%%%%
                \section{Optical data} \label{3c454:optical}
       %%%%%%%%%%%%%%%%%%%%%%%%%%%%%%%%%%%%%%%%%%%%%%%%%%%%%%%%%%%%%%%%%%%%
%
%%%%%--------------------------------------------
  \subsection{\webt{} Data Reduction and Analysis} \label{3c454:webt:analysis}
%%%%%--------------------------------------------
%
The Whole Earth Blazar Telescope (\webt)
\footnote{\texttt{http://www.oato.inaf.it/blazar/webt}, see e.g. 
\cite{Villata2004:WEBT:BLLac}.} has been monitoring 
\source{} since the exceptional 
2004--2005 outburst \citep{vil06,vil07,rai07,rai08a,rai08b}, 
covering also the period of the \agile{} observation.
We refer to \cite{rai08a} for a detailed presentation
and discussion of the radio, mm, near--IR, optical and \swi{}/UVOT data
collected, almost continously, during November 2007.

%%%%%--------------------------------------------
  \subsection{\rem{} Data Reduction and Analysis} \label{3c454:rem:analysis}
%%%%%--------------------------------------------
%
The photometric Optical and Near Infrared (NIR) observations were carried
out with
the Rapid Eye Mount \citep[REM,][]{Zerbi2004:REM}, a robotic telescope located
at the ESO Cerro La Silla observatory (Chile).
The REM  telescope has a Ritchey-Chretien configuration with a 60 cm f/2.2
primary and an overall f/8 focal ratio in a fast moving alt-azimuth mount
providing two stable Nasmyth focal stations. At one of the two foci the
telescope simultaneously feeds, by means of a dichroic, two cameras: REMIR
for the NIR \citep[][]{Conconi2004:REM}, and ROSS \citep[][]{Tosti2004:REM} 
for the optical.
Both the cameras have a field of view of $10'\times10'$ and imaging
capabilities with the usual NIR ($z$', $J$, $H$ and $K$) and 
Johnson-Cousins $V$, $R$, and $I$ filters.
All raw optical/NIR frames obtained with  REM Telescopes, were corrected for
dark, bias and flat field. Instrumental magnitudes were obtained via
aperture photometry using 
GAIA\footnote{\texttt{http://star-www.dur.ac.uk/$\sim$pdraper/gaia/gaia.html}}.
Calibration of the optical source magnitude was obtained by differential
photometry with respect to the comparison stars sequence reported by
\cite{Fiorucci1998:mags}  and \cite{Raiteri1998:mags}.  For the NIR
calibration we used the comparison sequence reported by 
\cite{Gonzales2001:mags}.
Both REMIR and ROSS instruments were used in order to obtain nearly 
simultaneous data and to study the spectral behavior \source{} at different
levels of flux.

%
       %%%%%%%%%%%%%%%%%%%%%%%%%%%%%%%%%%%%%%%%%%%%%%%%%%%%%%%%%%%%%%%%%%%%
            \section{Simultaneous data analysis} \label{3c454:sim:data}
\begin{figure}[!ht]
  \includegraphics[angle=0,scale=.9]{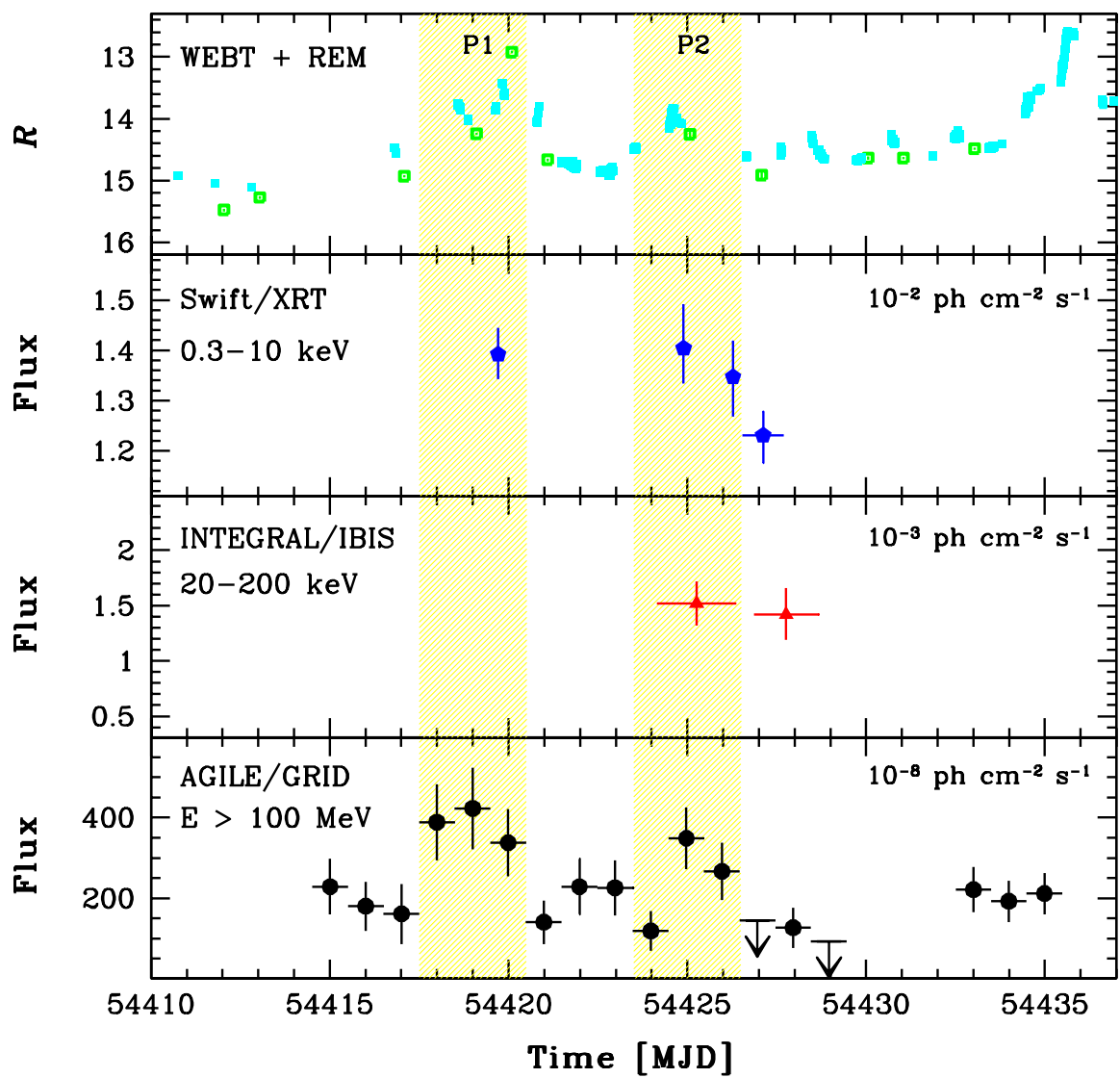}
  \caption{Simultaneous light curves acquired during the period
    2007 November 6--December 3. Black circles represent \agile{}/GRID
    data (30~MeV--50~GeV); red triangles represent \igr{}/IBIS data
    (20--200~keV); blue pentagons represent \swi{}/XRT data (0.3--10~keV);
    cyan--solid and green--open squares represent $R$-band WEBT and REM
    \citep{rai08a} data,
    respectively. The yellow areas mark the periods P1 and P2 during which
    we compute the simultaneous spectral energy distributions.
    [{\it See the electronic edition of the Journal for a color version of 
	this figure.}]
  }
  \label{3c454:fig:lcs}
\end{figure}

       %%%%%%%%%%%%%%%%%%%%%%%%%%%%%%%%%%%%%%%%%%%%%%%%%%%%%%%%%%%%%%%%%%%%
%
Figure~\ref{3c454:fig:lcs} shows the simultaneous light curves acquired
during the period 2007 November 6--December 3. Black circles represent 
\agile{}/GRID data (30~MeV--50~GeV); red triangles represent 
\igr{}/IBIS data (20--200~keV); blue pentagons represent \swi{}/XRT data 
(0.3--10~keV); cyan--solid and green--open squares represent 
$R$-band WEBT and REM
\citep{rai08a} data, respectively. The yellow areas mark 
the periods P1 and P2 during which we compute the simultaneous spectral 
energy distributions, and corresponding to higher \gray{} flux levels.
We note that during the period P1 the optical flux shows
intense variability, reaching a relative maximum on the last day
of the \gray day-by-day sampling.
Moreover, an optical flare as intense as the one on MJD$\sim$54420 
occurred at the end of the \agile{} observations (MJD$\sim$54435.5).
The three \gray data points show no particular flaring activity,
though the daily flux remained quite high, $\sim 200 \times 10^{-8}$\phcmsec.
A detailed discussion on the correlation between the optical and \gray data
during December 2007 will be presented in the forthcoming Paper II
(Donnarumma et al. 2008, in preparation).

\begin{figure}[ht]
\resizebox{\hsize}{!}{\includegraphics[angle=0]{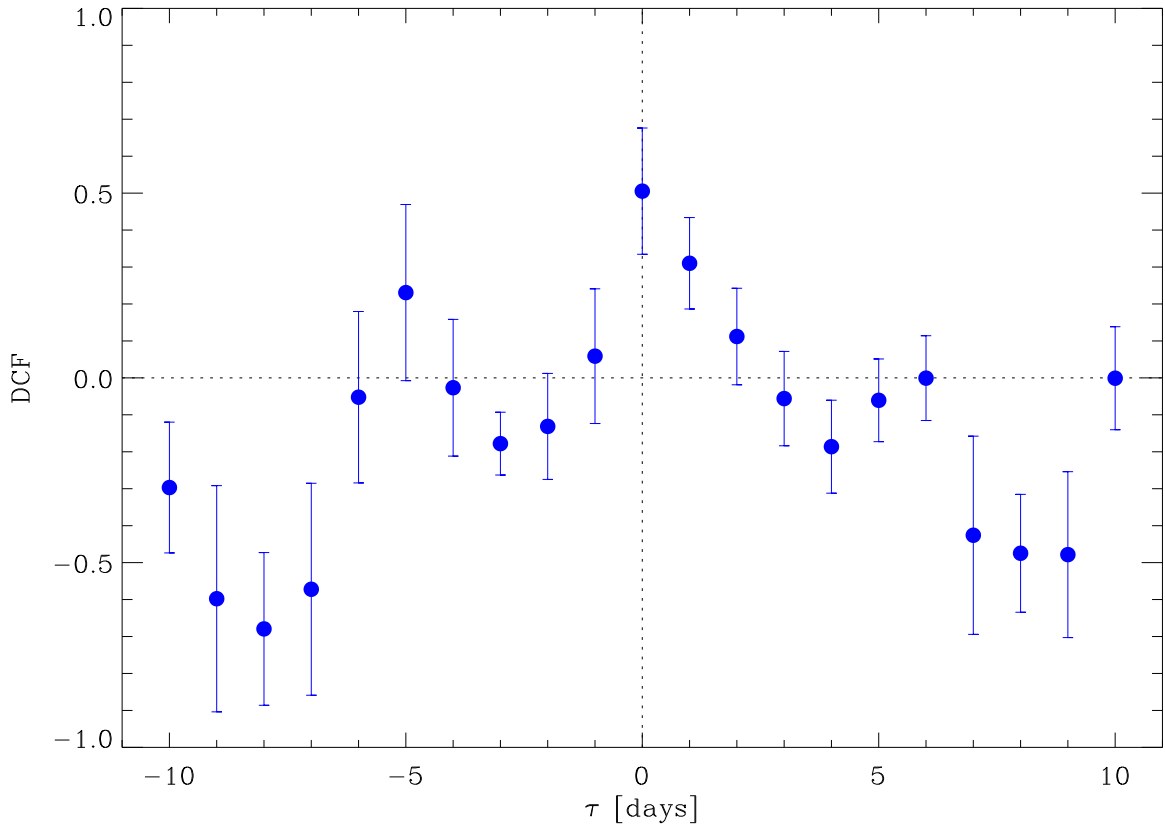}}
  \caption{Discrete correlation function between the \gray and optical
    fluxes. The optical data have previously been binned over 12 hours to smooth
    the intranight variations. The DCF peak suggests a mild correlation with no
    time delay.
    [{\it See the electronic edition of the Journal for a color version of 
      this figure.}]
  }
  \label{3c454:fig:webt:dcf}
\end{figure}

\begin{figure}[ht]
\resizebox{\hsize}{!}{\includegraphics[angle=0]{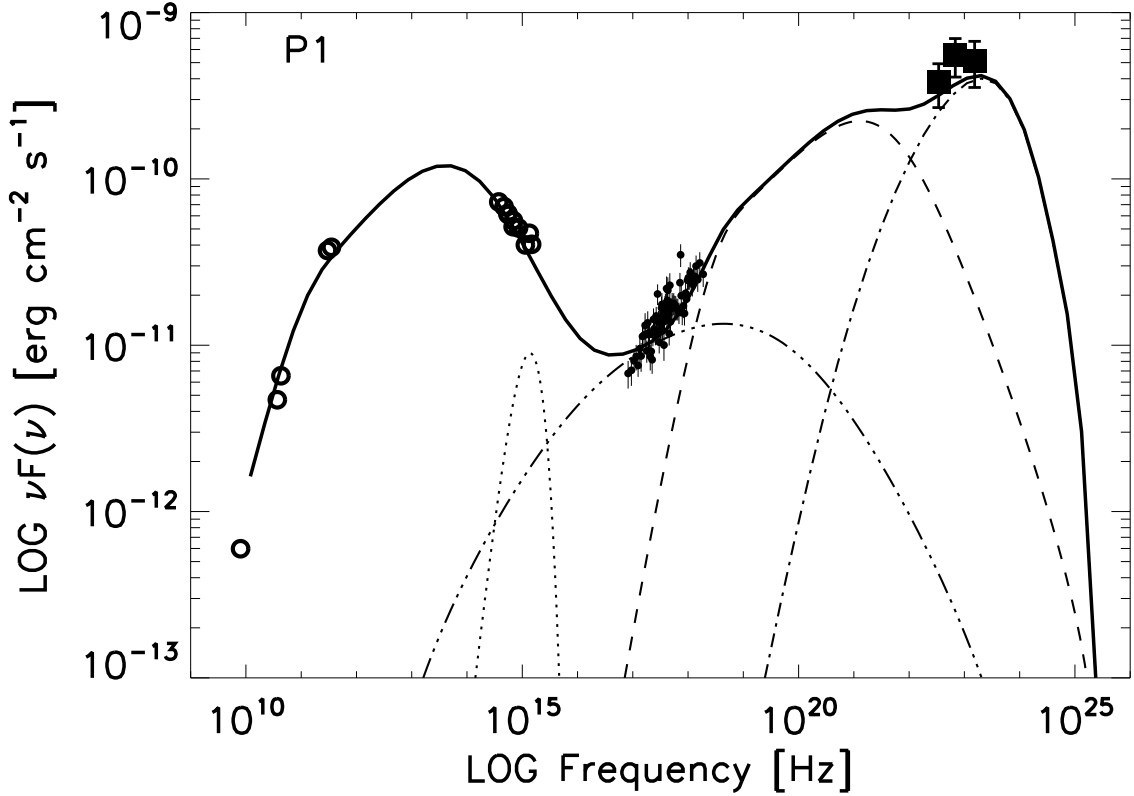}}
  \caption{Spectral energy distribution for the period P1, 
    MJD 54417.5--54420.5 (see Figure~\ref{3c454:fig:lcs}). Filled squares
    represent the \agile{}/GRID data in the
    energy range 100--1000~MeV; 
    small filled
    circles represent \swi{}/XRT data in the energy range 0.3--10~keV
    (segment 001); open symbols represent radio to UV data taken 
    from \cite{rai08a}, corresponding to MJD 54420.The dotted,
      dashed, dot--dashed, and triple--dot dashed lines represent 
      the accretion disk, the external
      Compton on the disk radiation, the external Compton on broad line
      region  radiation, and the SSC contributions, respectively.
  }
  \label{3c454:fig:sed:p1}
\end{figure}

We investigated the expected $\gamma$-optical flux correlation by means of
the discrete correlation function \citep[DCF; see][]{ede88,huf92,pet01}.
The result is shown in Figure~\ref{3c454:fig:webt:dcf}. 
The DCF peak occurred at $\tau=0$, and
its value is $\sim 0.5$. This indicates a moderate correlation, with no
significant time delay between the \gray and optical flux variations.
A minor peak at $\tau=-5$ days comes from establishing a connection between
the optical flare at MJD $\sim$ 54420 and the $\gamma$ high flux at MJD
$\sim$ 54425.
Figure~\ref{3c454:fig:sed:p1} shows the spectral energy distribution for 
the period P1, 
MJD 54417.5--54420.5 (see Figure~\ref{3c454:fig:lcs}). Filled squares
represent the \agile{}/GRID data in the
energy range 100--1000~MeV; 
small filled
circles represent \swi{}/XRT data in the energy range 0.3--10~keV 
(segment 001); open symbols represent radio to UV data taken from 
\cite{rai08a}, corresponding to MJD~54420, when all the WEBT
UBVRI bands were available, as well as \swi{}/UVOT data.

\begin{figure}[ht]
\resizebox{\hsize}{!}{\includegraphics[angle=0]{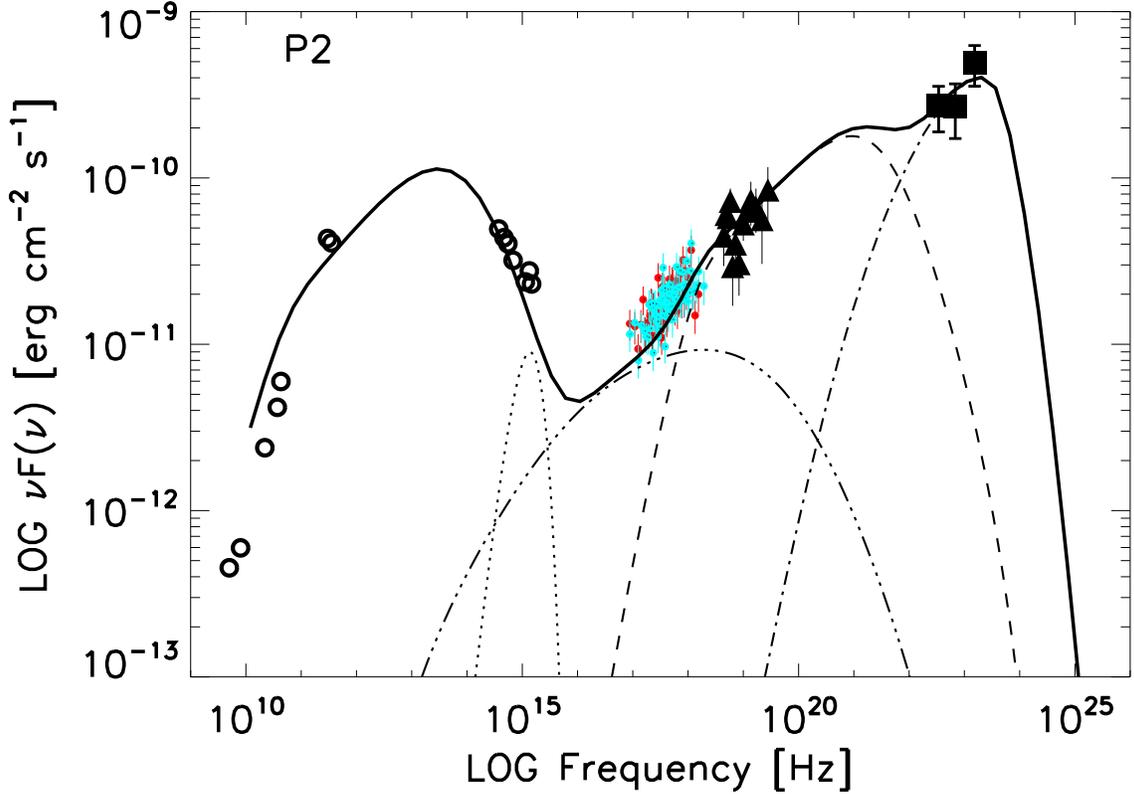}}
  \caption{Spectral energy distribution for 
    the period P2, 
    MJD 54423.5--54426.5 (see Figure~\ref{3c454:fig:lcs}). Filled squares
    represent the \agile{}/GRID data in the
    energy range 100--1000~MeV; filled triangles represent \igr/IBIS
    data in the energy range 20--200~keV (orbits 623$+$624); small filled
    circles represent \swi{}/XRT data in the energy range 0.3--10~keV
    (segments 003, 004, and 005);
    open symbols represent radio to UV data taken from 
    \cite{rai08a}, corresponding to MJD 54425. The dotted,
      dashed, dot--dashed, and the triple--dot dashed lines represent 
      the accretion disk, the external Compton on the disk radiation,
      the external Compton on broad line region  radiation,  and the SSC 
      contributions, respectively.
    [{\it See the electronic edition of the Journal for a color version of 
      this figure.}]
  }
  \label{3c454:fig:sed:p2}
\end{figure}

Figure~\ref{3c454:fig:sed:p2} shows the spectral energy distribution for 
the period P2, 
MJD 54423.5--54426.5 (see Figure~\ref{3c454:fig:lcs}). Filled squares
represent the \agile{}/GRID data in the
energy range 100--1000~MeV; filled triangles represent \igr/IBIS
data in the energy range 17--150~keV (orbits 623$+$624); small filled
circles represent \swi{}/XRT data in the energy range 0.3--10~keV
(segments 003, 004, and 005); open symbols represent radio to UV data 
taken from \cite{rai08a}, corresponding to MJD 54425.

A brief discussion of the modeling of both SEDs is presented 
 in the Section~\ref{3c454:disc}.
%
       %%%%%%%%%%%%%%%%%%%%%%%%%%%%%%%%%%%%%%%%%%%%%%%%%%%%%%%%%%%%%%%%%%%%
                       \section{Discussion} \label{3c454:disc}
       %%%%%%%%%%%%%%%%%%%%%%%%%%%%%%%%%%%%%%%%%%%%%%%%%%%%%%%%%%%%%%%%%%%%
%
The long--term \gray activity of \source{} is one of most
interesting discoveries achieved by \agile{} during its
first 6 months of observations.
The source was already detected in high state in July
2007 during a 1-week \agile{} ToO
triggered by an intense optical flare detected by the \webt{}. 
During that period, the
source reached its highest intensity level, with an average \gray
flux of $F_{\rm E\,>\,100\,MeV} = (280 \pm 40) \times 10^{-8}$\,\phcmsec.
In November 2007, \source{} showed prominent and
prolonged \gray activity, with flaring episodes on a timescale of
a few days and an average \gray flux of
$F_{\rm E\,>\,100\,MeV} = (170 \pm 13) \times 10^{-8}$\,\phcmsec. 
This renewed activity
triggered observations at different frequencies, allowing us to
obtain an almost {\it simultaneous} SED coverage on 14 decades
in energies.
We dubbed \source{} as {\it crazy diamond} because of its dramatic
variability at high energies revealed during the first half of the
\agile{} Observing Cycle--1. It has become clear that this source is
playing the same role for \agile{} as 3C~279 had for \egret{}. The
\source{} strong variability has also a clear signature
at lower frequencies. As reported in \cite{rai08a},
during the \agile{} observation, on MJD 54425 the source
showed an extremely variable behavior in the $R$ band,
with a brightening of 0.33 mag in 2.3 hours. In the same paper,
the authors report other episodes of fast variability with flux 
variations of several tenths of mag in a few hours.
Moreover, while in July 2007 \source{} exhibited its most intense
optical flare but with a very moderate degree of \gray flux variability 
on a day-by-day time--scale, during the November 2007 campaign 
(see Figure~\ref{3c454:fig:gammalc}), we note a significant \gray flux 
variability on short time--scales with
at least two distinct flaring episodes (P1 and P2).
It is interesting to note that in the optical band
\source{} also seems to display more rapid flares during the
fall--winter 2007 \webt{} campaign than those that occurred during the
July 2007 monitoring (\citealt{rai08b}).

\begin{figure}[ht]
\resizebox{\hsize}{!}{\includegraphics[angle=0]{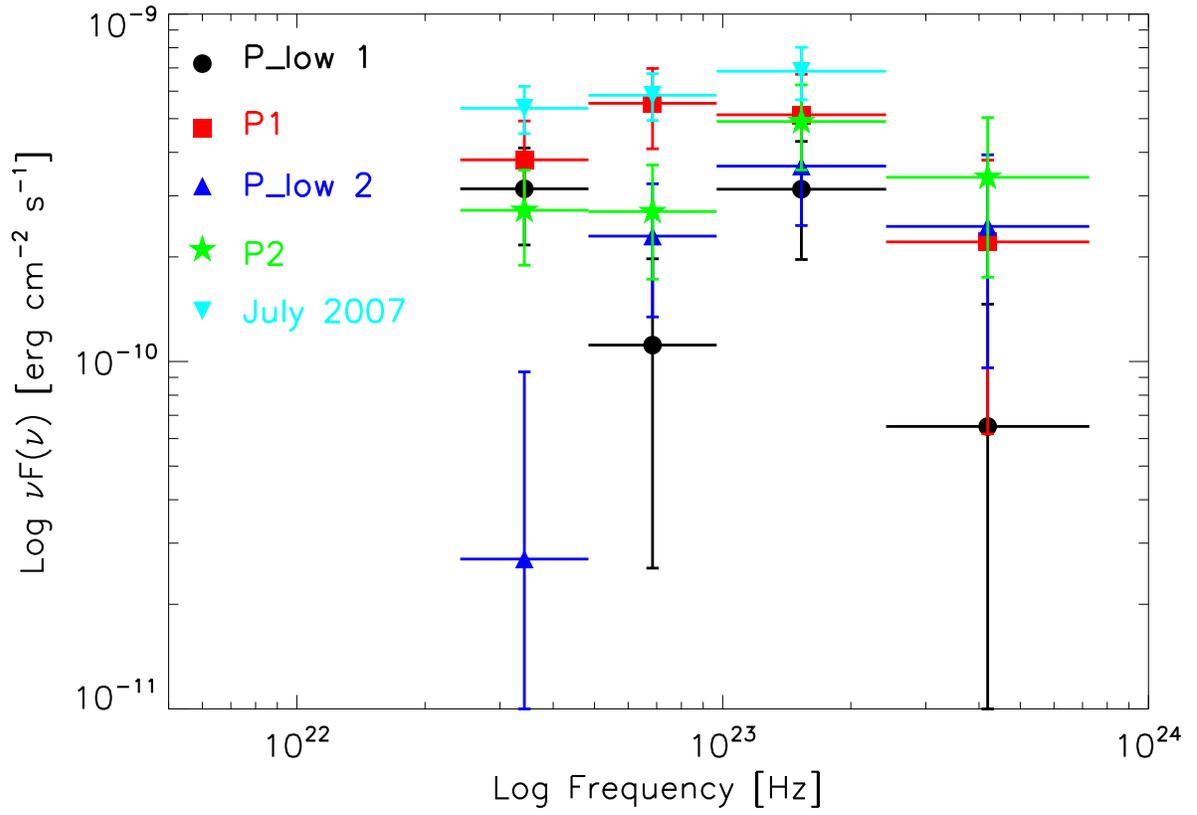}}
  \caption{\agile{}/GRID spectra for periods P1 (red squares),
    P2 (green stars), P1$\_$low1 (black circles), P2$\_$low2
    (blue upside triangles). The July 2007 spectrum is also
    shown (cyan upside down triangles).
  }
  \label{3c454:fig:spectra:PN}
\end{figure}

We compared the spectral properties of higher-state periods, P1 and P2,
with two lower-state periods, P$\_$low1 and P$\_$low2, chosen of the
same duration as P1 and P2 and corresponding to MJD 54414.5--54417.7 and
MJD 54420.5--54423.5, respectively.
Figure~\ref{3c454:fig:spectra:PN} shows the \agile{}/GRID spectra
for periods P1 (red square),
P2 (green star), P1$\_$low1 (black circle), P2$\_$low2
(blue upside triangle). The July 2007 spectrum is also
shown (cyan upside down triangle).
Although the statistics accumulated in only 4 days does not allow
us to obtain a robust fit of the data, Figure~\ref{3c454:fig:spectra:PN}
shows no clear spectral differences among different source intensity levels.

The correlation between the flux level and the spectral slope
was extensively studied by means of the analysis of the \egret{} data.
Recently, \cite{Nandikotkur2007:EGRET:slopes} have shown that
there is no homogeneous behavior among \egret{} blazars.
Although they consider {\it long--term} spectral dependence
on flux rather than {\it short--term} as in our case, our findings
are in agreement with their results on \source{}. Figure~3 in
\cite{Nandikotkur2007:EGRET:slopes} shows no spectral variation
despite a flux variation of about a factor of four.

Different emission mechanisms can be invoked to explain the \gray
emission and the different spectral states. In the leptonic
scenario, the low--frequency peak is interpreted as synchrotron
radiation from high--energy electrons in the relativistic jet,
while the high--energy peak can be produced by IC on different
{\it flavors} of seed photons. In the synchrotron
self--Compton [SSC] model (\citealt{Ghisellini1985:SSC},
\citealt{Bloom1996:SSC}) the seed photons come from the
jet itself. Alternatively, the seed photons can be those of the
accretion disk [external Compton scattering of direct disk
radiation, ECD, \cite{Dermer1992:ECD}]  or those of the broad--line
region (BLR) clouds [external Compton scattering from clouds, ECC,
\cite{Sikora1994:ECC}]. The target seed photons can also be those
produced by the infrared (IR) dust torus surrounding the nucleus
[external Compton scattering from IR dust, ERC(IR),
\cite{Sikora2002:ERCIR}].

\citealt{Blazejowski2000:MeV:blazars} showed that the ERC(IR) emission 
peak is at lower frequencies (soft \gray), more suitable for MeV blazars,
while our SEDs show a peak for the EC component in the GeV region
of the spectrum.
The average photon index during this November 2007 campaign
($\Gamma_{\rm AGILE} = 1.73 \pm 0.16$) is 
harder than the time--averaged one reported in 
\cite{Nandikotkur2007:EGRET:slopes}
($\Gamma_{\rm EGRET} = 2.22 \pm 0.06$) for \egret{}.
During intense \gray flares, the ECC and ECD processes play
a major role and the softness or the hardness of the resulting
spectrum is controlled by the dominant component, as
illustrated in \cite{Hartman2001:3C279:multiwave} for 3C~279.
In the case of \source{}, the ECC component seems to play a major
role, as we will show from the SED modeling.

We fit the SEDs for the P1 and P2 gamma-ray flaring episodes
by means of a one-zone leptonic model, considering the contributions 
from SSC and from external seed photons originating both from the
accretion disk and from the BLR. The emission along
the jet is assumed to be produced in
a spherical blob with comoving radius $R$ by
accelerated electrons characterized by a comoving broken power
law energy density distribution of the form,
\begin{equation}
n_{e}(\gamma)=\frac{K\gamma_{\rm b}^{-1}} 
{(\gamma/\gamma_{\rm b})^{\alpha_{\rm l}}+
(\gamma /\gamma_{\rm b})^{{\alpha}_{\rm h}}}\,,
\label{eq:ne_gamma}
\end{equation}
where $\gamma$ is the electron Lorentz factor assumed to vary
between $10<\gamma<1.5 \times 10^{4}$, $\alpha_{\rm l}$ and
$\alpha_{\rm h}$ are the pre-- and post--break electron distribution
spectral indices, respectively, and $\gamma_{\rm b}$ is the
break energy Lorentz factor. We assume that the blob
contains a random average magnetic field $B$ and that it
moves with a bulk Lorentz Factor
$\Gamma$ at an angle
$\Theta_{0}$  with respect to the line of sight. The
relativistic Doppler factor is then $\delta = [ \Gamma \,(1 - \beta
\, \cos{\Theta_{0}})]^{-1}$, where $\beta$ is the usual blob bulk
speed in units of the speed of light.

Our modelling of the \source{} high-energy emission is based
on an Inverse Compton  model with two main sources of external
target photons:
{\it (1)} an accretion disk characterized by a blackbody spectrum 
peaking in the UV with a bolometric luminosity  $L_{\rm d}$
for a IC-scattering blob at a distance $L$ 
from the central part of the disk;
{\it (2)} a Broad Line Region with a spectrum peaking in the $V$ band
and assumed to reprocess $10\%$ of the irradiating
continuum (\citealt{Tavecchio2008:blr:cloudy,rai07,rai08b}).

These two regions contribute to the ECD and the ECC,
respectively, and it is interesting to test the relative
importance of the two components that can be emitted by the
relativistic jet of \source{} under different conditions. A
complete theoretical analysis of the model, and of the interplay
among the different parameters is beyond the scope of this paper
and it will be presented in a forthcoming paper. We summarize here 
the main results of our best model characterized by an interesting 
set of physical parameters.

Table~\ref{3c454:tab:sed:param} shows the best-fit parameters
of our modelling of the flaring state SEDs corresponding to the
P1 and P2 phase of Figure~\ref{3c454:fig:lcs}. Our best fit parameters
values are: $B \sim 10$\,G, $\Gamma = 8.4$\,, 
$\Theta_{0} = 2.6^{\circ}$\,, and $r = 0.05$~pc\,
where $r$ is the distance between the accretion disk and
the emitting region.
In both Figures~\ref{3c454:fig:sed:p1} and~\ref{3c454:fig:sed:p2},
the dotted, dashed, and dot--dashed lines represent
the contributions of the accretion disk blackbody, the external
Compton on the disk radiation and the external Compton on
the broad line region radiation, respectively.

We note that during both the P1 and P2 episodes, the ECD
contribution can account for the soft and hard X-ray portion of
the spectrum, which show a moderate, if any, time
variability. However, we note that the ECD component alone
cannot account for the hardness of the \gray spectrum.
We therefore  argue that in the \agile{} energy band a
dominant contribution from ECC seems to provide a better fit of
the data during the gamma-ray flaring states P1 and P2. Moreover,
Table~\ref{3c454:tab:sed:param} shows that the data of both SEDs
can be fit by very similar model parameters. We note,
however, that  the high energy part of the electron energy
distribution appears to be softer during the  P2 episode as
compared to the electron distribution of the P1 flare.
\cite{Hartman2001:3C279:multiwave} find, for 3C~279, a 
relevant contribution of the SSC component in the 
X-rays - soft \gray bands. However, their average values
for $\gamma_{\rm 1}$ and $\gamma_{\rm 2}$ are much higher (a factor
of 5--10) than ours, resulting in an increase of the ratio 
${\rm SSC/Sync} \propto \gamma^{2}$\,. Thus, for the
Hartman et al. choice of parameters, the SSC contribution becomes
relevant at higher frequencies and of the same order of the ECD
contribution.

Our results can be compared with those obtained by
\cite{Jorstad2005:Lorentz}.
By means of total and polarized images obtained at the Very Long
Baseline Array (VLBA) at 7mm, they were able
to compute the global parameters of the source jet, estimating
$\langle \Gamma \rangle = 15.6 \pm 2.2$\,, $\langle \delta \rangle
= 24.6 \pm 4.5$\,, $\theta = (0.8 \pm 0.2)^{\circ}$\,, and $\langle
\Theta_{0} \rangle = (1.3 \pm 1.2)^{\circ}$\,, where $\langle \Gamma
\rangle$ and $\langle \delta \rangle$ are the average Lorentz and
Doppler factors, respectively, while $\theta$ and $\langle
\Theta_{0} \rangle$ are the intrinsic half opening angle of the
jet and the angle between the jet axis and the line of sight,
respectively.
We note, however, that the jet parameters derived so far
were obtained by means of data collected in earlier epochs with respect 
to our observations and refer to average values of different 
jet components.

The energetics of \source{} can be computed by estimating
the isotropic luminosity in the \gray band, $L_{\gamma}^{\rm iso}$, 
and comparing it with the Eddington, the bolometric, and the
particle injection luminosities.
For a given source with redshift $z$, the isotropic emitted
luminosity in the energy band $\epsilon$ is defined as
\begin{equation}
  \label{eq_lum}
  L(z)_{\epsilon} = \frac{4\pi F d_{\rm l}^{2}(z)}{(1+z)^{(1-\alpha)}} \,,
\end{equation}
where, in our case, $\epsilon$ is the \gray energy band
with ${E_{\rm min}}=100$\,MeV and  ${E_{\rm max}}=10$\,GeV,
$\alpha$ is the \gray energy spectral index, 
$\rm F(\nu) \propto \nu^{-\alpha}$ is the energy differential flux, 
$F=\int_{E_{\rm min}/h}^{E_{\rm max}/h}\,F(\nu)d\,\nu$ is the flux in the
\gray band,
and the luminosity distance is given by
\begin{equation}
  \label{dist_lum1}
  d_{\rm l}(z_1,z_2)= (1+z_2)^{2} \times \frac{c/H_0}{1+z_2}\int_{z_1}^{z_2}
  [E(z)]^{-1} dz\,,
\end{equation}
where $\rm z_1 = 0$, $\rm z_2 = z_{\rm src}$ and
\begin{equation}
  E(z)=\sqrt{\Omega_{\rm M} (1+z)^3
    +(1-\Omega_{\rm M}-\Omega_{\Lambda})(1+z)^2+\Omega_{\Lambda}}\,,
\end{equation}
where $H_0$ is the Hubble constant,  $\Omega_M$  and $\Omega_{\Lambda}$ are
the contribution of the matter and of the cosmological constant, respectively, to
the density parameter. Hereafter, we assume $H_0 = 70$\,km\,s$^{-1}$Mpc$^{-1}$,
 $\Omega_M = 0.3$,  and $\Omega_{\Lambda}=0.7$.
Using the observed average \gray flux, we obtain
$L_{\gamma}^{\rm iso} = 3.9\times10^{48}$\,\ergsec{}.

Moreover, from the values quoted in 
Table~\ref{3c454:tab:sed:param} and from Equation~\ref{eq:ne_gamma}
we can compute the particle
injection luminosity, $L_{\rm inj}$\,, obtaining :
\begin{equation}
L_{\rm inj} = \pi\,R^{2}\,\Gamma^{2}\,c\,\int[d\gamma\,\,m_{\rm e}\,c^{2}\gamma\,n(\gamma)] = 3 \times 10^{44}\,\ergsec\,.
\end{equation}

Assuming for \source{} a black hole mass $M_{\rm BH} =
4.4\times10^{9}$\,M$_{\odot}$ \citep{Gu2001:BH:masses}\,, we
obtain an Eddington luminosity of the order of $L_{\rm
Edd}=5.7\times10^{47}$\ergsec{} to be compared with the bolometric
luminosity $L_{\rm bol}=1.9\times10^{47}$\ergsec{} reported in
\cite{Woo2002:BH:masses}. 

We obtain, therefore, that the source energetic is comparable to
the value obtained by \cite{Tavecchio2007:jet:power} for
the power of the inner portion of jet, few $\times 10^{47}$\,\ergsec{}.
%
%

       %%%%%%%%%%%%%%%%%%%%%%%%%%%%%%%%%%%%%%%%%%%%%%%%%%%%%%%%%%%%%%%%%%%%
               \section{Conclusions} \label{3c454:conc}
       %%%%%%%%%%%%%%%%%%%%%%%%%%%%%%%%%%%%%%%%%%%%%%%%%%%%%%%%%%%%%%%%%%%%
%
%
The \agile{} mission is particularly suited to monitor a large number 
of potential \gray sources. The \agile{} pointings during the month 
of November 2007,  despite being centered approximately in the Cygnus 
region of the Galactic plane, revealed the very prominent \gray 
activity of the blazar \source{}. The \agile{} detection of this blazar 
prompted a series of important multiwavelength observations. The 
electromagnetic emission of 3C 454.3 could be determined with 
unprecedented coverage over 14 orders of magnitude in energy 
during a period that included a substantial fraction of the months
of November and December, 2007. 
Results of the AGILE data and related multifrequency campaign carried 
out in December, 2007, will be presented in a forthcoming paper.

We reported in this paper the main results of our \agile{} and 
associated multifrequency campaigns during the month of November, 2007.  
Our results can be summarized as follows:
\begin{enumerate}
\item The \gray emission from \source{} dominated the whole 
  extragalactic sky as monitored by AGILE during its first year 
  of scientific operations. 
\item Our \gray data show remarkable variability on a daily 
  timescale for a Flat Spectrum Radio Quasar.  
\item Emission in the optical range appears to be correlated with that 
  at \gray energies above 100 MeV.
\item Variability in the soft and hard X-ray range is less sensitive 
  to the short-timescale variations of the optical flux.
\item The average \gray spectrum during the whole campaign is 
  substantially harder than that reported in previous observations.
\item We determined the SEDs for episodes of relatively high \gray
  emission.
\item  Our results support the idea that the dominant emission mechanism 
  in \gray energy band is the inverse Compton scattering of external 
  photons from the BLR clouds scattering off the relativistic electrons 
  in the jet.
\end{enumerate}

       %%%%%%%%%%%%%%%%%%%%%%%%%%%%%%%%%%%%%%%%%%%%%%%%%%%%%%%%%%%%%%%%%%%%
\acknowledgements
We thank the Referee for his/her very prompt and constructive comments.
The AGILE Mission is funded by the Italian Space Agency (ASI) with
scientific and programmatic participation by the Italian Institute
of Astrophysics (INAF) and the Italian Institute of Nuclear
Physics (INFN). This investigation was carried out with partial
support under ASI contract N. {I/089/06/1}. This work is partly 
based on data taken and assembled by the 
WEBT collaboration and stored in the WEBT archive at the Osservatorio 
Astronomico di Torino - INAF (\texttt{http://www.to.astro.it/blazars/webt/}). 
We thank the \swi{} and \igr{} Teams for making these observations possible, 
particularly the duty scientists and science planners. 
MTF, AB and PU acknowledge ASI/INAF contract N. 023/05/0.
       %%%%%%%%%%%%%%%%%%%%%%%%%%%%%%%%%%%%%%%%%%%%%%%%%%%%%%%%%%%%%%%%%%%%

%
{\it Facilities:} \facility{AGILE}, \facility{Swift}, \facility{INTEGRAL},
\facility{WEBT}, \facility{REM}

%\bibliographystyle{apj}%%%%%%%%%%%%%%%%%%%%%%%%%%%%%%%%%%%%
%\bibliography{blazar}

\begin{thebibliography}{60}
\expandafter\ifx\csname natexlab\endcsname\relax\def\natexlab#1{#1}\fi

\bibitem[{{Aller} {et~al.}(1997){Aller}, {Marscher}, {Hartman}, {Aller},
  {Aller}, {Balonek}, {Begelman}, {Chiaberge}, {Clements}, {Collmar}, {de
  Francesco}, {Gear}, {Georganopoulos}, {Ghisellini}, {Glass},
  {Gonzalez-Perez}, {Heinimaki}, {Herter}, {Hooper}, {Hughes}, {Johnson},
  {Katajainen}, {Kidger}, {Kraus}, {Lanteri}, {Lawrence}, {Lichti}, {Lin},
  {Madejski}, {McNaron-Brown}, {Moore}, {Mukherjee}, {Nair}, {Nilsson},
  {Peila}, {Pierkowski}, {Pohl}, {Pursimo}, {Raiteri}, {Reich}, {Robson},
  {Sillanpaa}, {Sikora}, {Smith}, {Steppe}, {Stevens}, {Takalo}, {Terasranta},
  {Tornikoski}, {Valtaoja}, {von Montigny}, {Villata}, {Wagner}, {Wichmann}, \&
  {Witzel}}]{Aller1997:3C454_EGRET}
{Aller}, M.~F. {et~al.} 1997, in American Institute of Physics Conference
  Series, Vol. 410, Proceedings of the Fourth Compton Symposium, ed. C.~D.
  {Dermer}, M.~S. {Strickman}, \& J.~D. {Kurfess}, 1423

\bibitem[{{Barthelmy} {et~al.}(2005){Barthelmy}, {Barbier}, {Cummings},
  {Fenimore}, {Gehrels}, {Hullinger}, {Krimm}, {Markwardt}, {Palmer},
  {Parsons}, {Sato}, {Suzuki}, {Takahashi}, {Tashiro}, \&
  {Tueller}}]{Barthelmy2005SSRv:bat}
{Barthelmy}, S.~D. {et~al.} 2005, Space Science Reviews, 120, 143

\bibitem[{{B{\l}a{\.z}ejowski} {et~al.}(2000){B{\l}a{\.z}ejowski}, {Sikora},
  {Moderski}, \& {Madejski}}]{Blazejowski2000:MeV:blazars}
{B{\l}a{\.z}ejowski}, M., {Sikora}, M., {Moderski}, R., \& {Madejski}, G.~M.
  2000, \apj, 545, 107

\bibitem[{{Bloom} \& {Marscher}(1996)}]{Bloom1996:SSC}
{Bloom}, S.~D., \& {Marscher}, A.~P. 1996, \apj, 461, 657

\bibitem[{{Burrows} {et~al.}(2005){Burrows}, {Hill}, {Nousek}, {Kennea},
  {Wells}, {Osborne}, {Abbey}, {Beardmore}, {Mukerjee}, {Short}, {Chincarini},
  {Campana}, {Citterio}, {Moretti}, {Pagani}, {Tagliaferri}, {Giommi},
  {Capalbi}, {Tamburelli}, {Angelini}, {Cusumano}, {Br{\"a}uninger}, {Burkert},
  \& {Hartner}}]{Burrows2005:grades}
{Burrows}, D.~N. {et~al.} 2005, Space Science Reviews, 120, 165

\bibitem[{{Celotti} \& {Ghisellini}(2008)}]{Celotti2008:blazar:jet}
{Celotti}, A., \& {Ghisellini}, G. 2008, \mnras, 385, 283

\bibitem[{{Chen} {et~al.}(2007){Chen}, {Vercellone}, {Giuliani}, {Mereghetti},
  {Pellizzoni}, {Perotti}, {Fornari}, {Fiorini}, {Caraveo}, {Zambra},
  {Bulgarelli}, {Gianotti}, {Trifoglio}, {Cocco}, {Labanti}, {Fuschino},
  {Marisaldi}, {Galli}, {Tavani}, {Pucella}, {D'Ammando}, {Vittorini}, {Costa},
  {Feroci}, {Donnarumma}, {Pacciani}, {Monte}, {Lazzarotto}, {Soffitta},
  {Evangelista}, {Lapshov}, {Rapisarda}, {Argan}, {Trois}, {Paris},
  {Barbiellini}, {Longo}, {Picozza}, {Morselli}, {Prest}, {Vallazza}, {Lipari},
  {Zanello}, {Mauri}, {Giommi}, {Pittori}, {Antonelli}, {Gasparrini}, {Cutini},
  {Verrecchia}, \& {Salotti}}]{Chen2007:ATel3C454}
{Chen}, A. {et~al.} 2007, The Astronomer's Telegram, 1278, 1

\bibitem[{{Conconi} {et~al.}(2004){Conconi}, {Cunniffe}, {D'Alessio},
  {Calzoletti}, {Jordan}, {Mazzoleni}, {Melandri}, {Molinari}, {Testa},
  {Vitali}, {Zerbi}, {Chincarini}, {Covino}, {Ghisellini}, {Rodono}, {Tosti},
  {Antonelli}, {Cutispoto}, {Nicastro}, \& {Palazzi}}]{Conconi2004:REM}
{Conconi}, P. {et~al.} 2004, in Presented at the Society of Photo-Optical
  Instrumentation Engineers (SPIE) Conference, Vol. 5492, Ground-based
  Instrumentation for Astronomy. Edited by Alan F. M. Moorwood and Iye
  Masanori. Proceedings of the SPIE, Volume 5492, pp. 1602-1612 (2004)., ed.
  A.~F.~M. {Moorwood} \& M.~{Iye}, 1602--1612

\bibitem[{{Dermer} {et~al.}(1992){Dermer}, {Schlickeiser}, \&
  {Mastichiadis}}]{Dermer1992:ECD}
{Dermer}, C.~D., {Schlickeiser}, R., \& {Mastichiadis}, A. 1992, \aap, 256, L27

\bibitem[{{Edelson} \& {Krolik}(1988)}]{ede88}
{Edelson}, R.~A., \& {Krolik}, J.~H. 1988, \apj, 333, 646

\bibitem[{{Feroci} {et~al.}(2007){Feroci}, {Costa}, {Soffitta}, {Del Monte},
  {di Persio}, {Donnarumma}, {Evangelista}, {Frutti}, {Lapshov}, {Lazzarotto},
  {Mastropietro}, {Morelli}, {Pacciani}, {Porrovecchio}, {Rapisarda}, {Rubini},
  {Tavani}, \& {Argan}}]{Feroci2007:agile_sa}
{Feroci}, M. {et~al.} 2007, Nuclear Instruments and Methods in Physics Research
  A, 581, 728

\bibitem[{{Fiorucci} {et~al.}(1998){Fiorucci}, {Tosti}, \&
  {Rizzi}}]{Fiorucci1998:mags}
{Fiorucci}, M., {Tosti}, G., \& {Rizzi}, N. 1998, \pasp, 110, 105

\bibitem[{{Fuhrmann} {et~al.}(2006){Fuhrmann}, {Cucchiara}, {Marchili},
  {Tosti}, {Nucciarelli}, {Ciprini}, {Molinari}, {Chincarini}, {Zerbi},
  {Covino}, {Pian}, {Meurs}, {Testa}, {Vitali}, {Antonelli}, {Conconi},
  {Cutispoto}, {Malaspina}, {Nicastro}, {Palazzi}, \&
  {Ward}}]{Fuhrmann2006:3c454:opt}
{Fuhrmann}, L. {et~al.} 2006, \aap, 445, L1

\bibitem[{{Gehrels} {et~al.}(2004){Gehrels}, {Chincarini}, {Giommi}, {Mason},
  {Nousek}, {Wells}, {White}, {Barthelmy}, {Burrows}, {Cominsky}, {Hurley},
  {Marshall}, {M{\'e}sz{\'a}ros}, {Roming}, {Angelini}, {Barbier}, {Belloni},
  {Campana}, {Caraveo}, {Chester}, {Citterio}, {Cline}, {Cropper}, {Cummings},
  {Dean}, {Feigelson}, {Fenimore}, {Frail}, {Fruchter}, {Garmire}, {Gendreau},
  {Ghisellini}, {Greiner}, {Hill}, {Hunsberger}, {Krimm}, {Kulkarni}, {Kumar},
  {Lebrun}, {Lloyd-Ronning}, {Markwardt}, {Mattson}, {Mushotzky}, {Norris},
  {Osborne}, {Paczynski}, {Palmer}, {Park}, {Parsons}, {Paul}, {Rees},
  {Reynolds}, {Rhoads}, {Sasseen}, {Schaefer}, {Short}, {Smale}, {Smith},
  {Stella}, {Tagliaferri}, {Takahashi}, {Tashiro}, {Townsley}, {Tueller},
  {Turner}, {Vietri}, {Voges}, {Ward}, {Willingale}, {Zerbi}, \&
  {Zhang}}]{Gehrels2004:swift}
{Gehrels}, N. {et~al.} 2004, \apj, 611, 1005

\bibitem[{{Ghisellini} {et~al.}(2007){Ghisellini}, {Foschini}, {Tavecchio}, \&
  {Pian}}]{Ghisellini2007:3C454:SED}
{Ghisellini}, G., {Foschini}, L., {Tavecchio}, F., \& {Pian}, E. 2007, \mnras,
  382, L82

\bibitem[{{Ghisellini} {et~al.}(1985){Ghisellini}, {Maraschi}, \&
  {Treves}}]{Ghisellini1985:SSC}
{Ghisellini}, G., {Maraschi}, L., \& {Treves}, A. 1985, \aap, 146, 204

\bibitem[{{Giommi} {et~al.}(2006){Giommi}, {Blustin}, {Capalbi},
  {Colafrancesco}, {Cucchiara}, {Fuhrmann}, {Krimm}, {Marchili}, {Massaro},
  {Perri}, {Tagliaferri}, {Tosti}, {Tramacere}, {Burrows}, {Chincarini},
  {Falcone}, {Gehrels}, {Kennea}, \& {Sambruna}}]{Giommi2006:3C454_Swift}
{Giommi}, P. {et~al.} 2006, \aap, 456, 911

\bibitem[{{Giuliani} {et~al.}(2004){Giuliani}, {Chen}, {Mereghetti},
  {Pellizzoni}, {Tavani}, \& {Vercellone}}]{Giuliani2004:diff_model}
{Giuliani}, A., {Chen}, A., {Mereghetti}, S., {Pellizzoni}, A., {Tavani}, M.,
  \& {Vercellone}, S. 2004, Mem.\ SAIt Suppl., 5, 135

\bibitem[{{Gonz{\'a}lez-P{\'e}rez} {et~al.}(2001){Gonz{\'a}lez-P{\'e}rez},
  {Kidger}, \& {Mart{\'{\i}}n-Luis}}]{Gonzales2001:mags}
{Gonz{\'a}lez-P{\'e}rez}, J.~N., {Kidger}, M.~R., \& {Mart{\'{\i}}n-Luis}, F.
  2001, \aj, 122, 2055

\bibitem[{{Gu} {et~al.}(2001){Gu}, {Cao}, \& {Jiang}}]{Gu2001:BH:masses}
{Gu}, M., {Cao}, X., \& {Jiang}, D.~R. 2001, \mnras, 327, 1111

\bibitem[{{Hartman} {et~al.}(1999){Hartman}, {Bertsch}, {Bloom}, {Chen},
  {Deines-Jones}, {Esposito}, {Fichtel}, {Friedlander}, {Hunter}, {McDonald},
  {Sreekumar}, {Thompson}, {Jones}, {Lin}, {Michelson}, {Nolan}, {Tompkins},
  {Kanbach}, {Mayer-Hasselwander}, {M{\"u}cke}, {Pohl}, {Reimer}, {Kniffen},
  {Schneid}, {von Montigny}, {Mukherjee}, \& {Dingus}}]{Hartman1999:3eg}
{Hartman}, R.~C. {et~al.} 1999, \apjs, 123, 79

\bibitem[{{Hartman} {et~al.}(1993){Hartman}, {Bertsch}, {Dingus}, {Fichtel},
  {Hunter}, {Kanbach}, {Kniffen}, {Lin}, {Mattox}, {Mayer-Hasselwander},
  {Michelson}, {von Montigny}, {Nolan}, {Piner}, {Schneid}, {Sreekumar}, \&
  {Thompson}}]{Hartman1993:3C454_EGRET}
---. 1993, \apjl, 407, L41

\bibitem[{{Hartman} {et~al.}(1992){Hartman}, {Bertsch}, {Fichtel}, {Hunter},
  {Kwok}, {Mattox}, {Sreekumar}, {Thompson}, {Kniffen}, {Lin}, {Michelson},
  {Nolan}, {Schneid}, {Kanbach}, {Mayer-Hasselwander}, {von Montigny},
  {Pinkau}, {Rothermel}, \& {Sommer}}]{Hartman1992:3C454iauc}
---. 1992, \iaucirc, 5477, 2

\bibitem[{{Hartman} {et~al.}(2001){Hartman}, {B{\"o}ttcher}, {Aldering},
  {Aller}, {Aller}, {Backman}, {Balonek}, {Bertsch}, {Bloom}, {Bock},
  {Boltwood}, {Carini}, {Collmar}, {De Francesco}, {Ferrara}, {Freudling},
  {Gear}, {Hall}, {Heidt}, {Hughes}, {Hunter}, {Jogee}, {Johnson}, {Kanbach},
  {Katajainen}, {Kidger}, {Kii}, {Koskimies}, {Kraus}, {Kubo}, {Kurtanidze},
  {Lanteri}, {Lawson}, {Lin}, {Lisenfeld}, {Madejski}, {Makino}, {Maraschi},
  {Marscher}, {McFarland}, {McHardy}, {Miller}, {Nikolashvili}, {Nilsson},
  {Noble}, {Nucciarelli}, {Ostorero}, {Pian}, {Pursimo}, {Raiteri}, {Reich},
  {Rekola}, {Richter}, {Robson}, {Sadun}, {Savolainen}, {Sillanp{\"a}{\"a}},
  {Smale}, {Sobrito}, {Sreekumar}, {Stevens}, {Takalo}, {Tavecchio},
  {Ter{\"a}sranta}, {Thompson}, {Tornikoski}, {Tosti}, {Ungerechts}, {Urry},
  {Valtaoja}, {Villata}, {Wagner}, {Wehrle}, \&
  {Wilson}}]{Hartman2001:3C279:multiwave}
---. 2001, \apj, 553, 683

\bibitem[{{Hufnagel} \& {Bregman}(1992)}]{huf92}
{Hufnagel}, B.~R., \& {Bregman}, J.~N. 1992, \apj, 386, 473

\bibitem[{{Jorstad} {et~al.}(2005){Jorstad}, {Marscher}, {Lister}, {Stirling},
  {Cawthorne}, {Gear}, {G{\'o}mez}, {Stevens}, {Smith}, {Forster}, \&
  {Robson}}]{Jorstad2005:Lorentz}
{Jorstad}, S.~G. {et~al.} 2005, \aj, 130, 1418

\bibitem[{{Kalberla} {et~al.}(2005){Kalberla}, {Burton}, {Hartmann}, {Arnal},
  {Bajaja}, {Morras}, \& {P{\"o}ppel}}]{Kalberla2005:nh}
{Kalberla}, P.~M.~W., {Burton}, W.~B., {Hartmann}, D., {Arnal}, E.~M.,
  {Bajaja}, E., {Morras}, R., \& {P{\"o}ppel}, W.~G.~L. 2005, \aap, 440, 775

\bibitem[{{Labanti} {et~al.}(2006){Labanti}, {Marisaldi}, {Fuschino}, {Galli},
  {Argan}, {Bulgarelli}, {Costa}, {Di Cocco}, {Gianotti}, {Tavani}, \&
  {Trifoglio}}]{Labanti2006:agile_mcal}
{Labanti}, C. {et~al.} 2006, in Proc. SPIE, 6266, 62663Q

\bibitem[{{Mattox} {et~al.}(1993){Mattox}, {Bertsch}, {Chiang}, {Dingus},
  {Fichtel}, {Hartman}, {Hunter}, {Kanbach}, {Kniffen}, {Kwok}, {Lin},
  {Mayer-Hasselwander}, {Michelson}, {von Montigny}, {Nolan}, {Pinkau},
  {Schneid}, {Sreekumar}, \& {Thompson}}]{Mattox1993:1633}
{Mattox}, J.~R. {et~al.} 1993, \apj, 410, 609

\bibitem[{{McLaughlin} {et~al.}(1996){McLaughlin}, {Mattox}, {Cordes}, \&
  {Thompson}}]{McLaughlin1996:v:index}
{McLaughlin}, M.~A., {Mattox}, J.~R., {Cordes}, J.~M., \& {Thompson}, D.~J.
  1996, \apj, 473, 763

\bibitem[{{M{\"u}cke} \& {Protheroe}(2001)}]{Mucke2001:hadronic}
{M{\"u}cke}, A., \& {Protheroe}, R.~J. 2001, Astroparticle Physics, 15, 121

\bibitem[{{M{\"u}cke} {et~al.}(2003){M{\"u}cke}, {Protheroe}, {Engel},
  {Rachen}, \& {Stanev}}]{Mucke2003:hadronic}
{M{\"u}cke}, A., {Protheroe}, R.~J., {Engel}, R., {Rachen}, J.~P., \& {Stanev},
  T. 2003, Astroparticle Physics, 18, 593

\bibitem[{{Nandikotkur} {et~al.}(2007){Nandikotkur}, {Jahoda}, {Hartman},
  {Mukherjee}, {Sreekumar}, {B{\"o}ttcher}, {Sambruna}, \&
  {Swank}}]{Nandikotkur2007:EGRET:slopes}
{Nandikotkur}, G., {Jahoda}, K.~M., {Hartman}, R.~C., {Mukherjee}, R.,
  {Sreekumar}, P., {B{\"o}ttcher}, M., {Sambruna}, R.~M., \& {Swank}, J.~H.
  2007, \apj, 657, 706

\bibitem[{{Perotti} {et~al.}(2006){Perotti}, {Fiorini}, {Incorvaia},
  {Mattaini}, \& {Sant'Ambrogio}}]{Perotti2006:agile_ac}
{Perotti}, F., {Fiorini}, M., {Incorvaia}, S., {Mattaini}, E., \&
  {Sant'Ambrogio}, E. 2006, Nuclear Instruments and Methods in Physics Research
  A, 556, 228

\bibitem[{{Peterson}(2001)}]{pet01}
{Peterson}, B.~M. 2001, in Advanced Lectures on the Starburst-AGN Connection,
  ed. I.~{Aretxaga}, D.~{Kunth}, \& R.~{M{\'u}jica} (Singapore: World
  Scientific), 3

\bibitem[{{Pian} {et~al.}(2006){Pian}, {Foschini}, {Beckmann}, {Soldi},
  {T{\"u}rler}, {Gehrels}, {Ghisellini}, {Giommi}, {Maraschi}, {Pursimo},
  {Raiteri}, {Tagliaferri}, {Tornikoski}, {Tosti}, {Treves}, {Villata}, {Barr},
  {Courvoisier}, {di Cocco}, {Hudec}, {Fuhrmann}, {Malaguti}, {Persic},
  {Tavecchio}, \& {Walter}}]{Pian2006:3C454_Integral}
{Pian}, E. {et~al.} 2006, \aap, 449, L21

\bibitem[{{Prest} {et~al.}(2003){Prest}, {Barbiellini}, {Bordignon}, {Fedel},
  {Liello}, {Longo}, {Pontoni}, \& {Vallazza}}]{Prest2003:agile_st}
{Prest}, M., {Barbiellini}, G., {Bordignon}, G., {Fedel}, G., {Liello}, F.,
  {Longo}, F., {Pontoni}, C., \& {Vallazza}, E. 2003, Nuclear Instruments and
  Methods in Physics Research A, 501, 280

\bibitem[{{Pucella} {et~al.}(2007){Pucella}, {Tavani}, {D'Ammando},
  {Vittorini}, {Costa}, {Feroci}, {Donnarumma}, {Pacciani}, {Monte},
  {Lazzarotto}, {Soffitta}, {Evangelista}, {Lapshov}, {Rapisarda}, {Argan},
  {Trois}, {Paris}, {Vercellone}, {Chen}, {Giuliani}, {Mereghetti},
  {Pellizzoni}, {Perotti}, {Fornari}, {Fiorini}, {Caraveo}, {Zambra},
  {Bulgarelli}, {Gianotti}, {Trifoglio}, {Cocco}, {Labanti}, {Fuschino},
  {Marisaldi}, {Galli}, {Barbiellini}, {Longo}, {Picozza}, {Morselli}, {Prest},
  {Vallazza}, {Lipari}, {Zanello}, {Mauri}, {Giommi}, {Pittori}, {Antonelli},
  {Gasparrini}, {Cutini}, {Verrecchia}, \& {Salotti}}]{Pucella2007:ATel3C454}
{Pucella}, G. {et~al.} 2007, The Astronomer's Telegram, 1300, 1

\bibitem[{{Raiteri} {et~al.}(2008{\natexlab{a}}){Raiteri}, {Villata}, {Chen},
  {Hsiao}, {Kurtanidze}, {Nilsson}, {Larionov}, {Gurwell}, {Agudo}, {Aller},
  {Aller}, {Angelakis}, {Arkharov}, {Bach}, {B{\"o}ttcher}, {Buemi},
  {Calcidese}, {Charlot}, {D'Ammando}, {Donnarumma}, {Forn{\'e}}, {Frasca},
  {Fuhrmann}, {G{\'o}mez}, {Hagen-Thorn}, {Jorstad}, {Kimeridze}, {Krichbaum},
  {L{\"a}hteenm{\"a}ki}, {Lanteri}, {Latev}, {Le Campion}, {Lee}, {Leto},
  {Lin}, {Marchili}, {Marilli}, {Marscher}, {Nesci}, {Nieppola},
  {Nikolashvili}, {Ohlert}, {Ovcharov}, {Principe}, {Pursimo}, {Ragozzine},
  {Sadun}, {Sigua}, {Smart}, {Strigachev}, {Takalo}, {Tavani}, {Thum},
  {Tornikoski}, {Trigilio}, {Uckert}, {Umana}, {Valcheva}, {Vercellone},
  {Volvach}, \& {Wiesemeyer}}]{rai08a}
{Raiteri}, C.~M. {et~al.} 2008{\natexlab{a}}, \aap, 485, L17

\bibitem[{{Raiteri} {et~al.}(1998){Raiteri}, {Villata}, {Lanteri}, {Cavallone},
  \& {Sobrito}}]{Raiteri1998:mags}
{Raiteri}, C.~M., {Villata}, M., {Lanteri}, L., {Cavallone}, M., \& {Sobrito},
  G. 1998, \aaps, 130, 495

\bibitem[{{Raiteri} {et~al.}(2008{\natexlab{b}}){Raiteri}, {Villata},
  {Larionov}, {Chen}, {Kurtanidze}, {Gurwell}, \& {et al.}}]{rai08b}
{Raiteri}, C.~M., {Villata}, M., {Larionov}, V.~M., {Chen}, W.~P.,
  {Kurtanidze}, O.~M., {Gurwell}, \& {et al.} 2008{\natexlab{b}}, \aap,
  submitted

\bibitem[{{Raiteri} {et~al.}(2007){Raiteri}, {Villata}, {Larionov}, {Pursimo},
  {Ibrahimov}, {Nilsson}, {Aller}, {Kurtanidze}, {Foschini}, {Ohlert},
  {Papadakis}, {Sumitomo}, {Volvach}, {Aller}, {Arkharov}, {Bach}, {Berdyugin},
  {B{\"o}ttcher}, {Buemi}, {Calcidese}, {Charlot}, {Delgado S{\'a}nchez}, {di
  Paola}, {Djupvik}, {Dolci}, {Efimova}, {Fan}, {Forn{\'e}}, {Gomez}, {Gupta},
  {Hagen-Thorn}, {Hooks}, {Hovatta}, {Ishii}, {Kamada}, {Konstantinova},
  {Kopatskaya}, {Kovalev}, {Kovalev}, {L{\"a}hteenm{\"a}ki}, {Lanteri}, {Le
  Campion}, {Lee}, {Leto}, {Lin}, {Lindfors}, {Mingaliev}, {Mizoguchi},
  {Nicastro}, {Nikolashvili}, {Nishiyama}, {{\"O}stman}, {Ovcharov},
  {P{\"a}{\"a}kk{\"o}nen}, {Pasanen}, {Pian}, {Rector}, {Ros}, {Sadakane},
  {Selj}, {Semkov}, {Sharapov}, {Somero}, {Stanev}, {Strigachev}, {Takalo},
  {Tanaka}, {Tavani}, {Torniainen}, {Tornikoski}, {Trigilio}, {Umana},
  {Vercellone}, {Valcheva}, {Volvach}, \& {Yamanaka}}]{rai07}
{Raiteri}, C.~M. {et~al.} 2007, \aap, 473, 819

\bibitem[{{Roming} {et~al.}(2005){Roming}, {Kennedy}, {Mason}, {Nousek}, {Ahr},
  {Bingham}, {Broos}, {Carter}, {Hancock}, {Huckle}, {Hunsberger}, {Kawakami},
  {Killough}, {Koch}, {McLelland}, {Smith}, {Smith}, {Soto}, {Boyd},
  {Breeveld}, {Holland}, {Ivanushkina}, {Pryzby}, {Still}, \&
  {Stock}}]{Roming2005SSRv:UVOT}
{Roming}, P.~W.~A. {et~al.} 2005, Space Science Reviews, 120, 95

\bibitem[{{Senziani} {et~al.}(2007){Senziani}, {Novara}, {de Luca}, {Caraveo},
  {Belloni}, \& {Bignami}}]{Senziani2007:bat:survey}
{Senziani}, F., {Novara}, G., {de Luca}, A., {Caraveo}, P.~A., {Belloni}, T.,
  \& {Bignami}, G.~F. 2007, \aap, 476, 1297

\bibitem[{{Sikora} {et~al.}(1994){Sikora}, {Begelman}, \&
  {Rees}}]{Sikora1994:ECC}
{Sikora}, M., {Begelman}, M.~C., \& {Rees}, M.~J. 1994, \apj, 421, 153

\bibitem[{{Sikora} {et~al.}(2002){Sikora}, {B{\l}a{\.z}ejowski}, {Moderski}, \&
  {Madejski}}]{Sikora2002:ERCIR}
{Sikora}, M., {B{\l}a{\.z}ejowski}, M., {Moderski}, R., \& {Madejski}, G.~M.
  2002, \apj, 577, 78

\bibitem[{{Tavani} {et~al.}(2008{\natexlab{a}}){Tavani}, {Barbiellini},
  {Argan}, {Bulgarelli}, {Caraveo}, {Chen}, {Cocco}, {Costa}, {de Paris}, {Del
  Monte}, {di Cocco}, {Donnarumma}, {Feroci}, {Fiorini}, {Froysland},
  {Fuschino}, {Galli}, {Gianotti}, {Giuliani}, {Evangelista}, {Labanti},
  {Lapshov}, {Lazzarotto}, {Lipari}, {Longo}, {Marisaldi}, {Mastropietro},
  {Mauri}, {Mereghetti}, {Morelli}, {Morselli}, {Pacciani}, {Pellizzoni},
  {Perotti}, {Picozza}, {Pontoni}, {Porrovecchio}, {Prest}, {Pucella},
  {Rapisarda}, {Rossi}, {Rubini}, {Soffitta}, {Trifoglio}, {Trois}, {Vallazza},
  {Vercellone}, {Zambra}, {Zanello}, {Giommi}, {Antonelli}, \&
  {Pittori}}]{Tavani2008_agile_nima}
{Tavani}, M. {et~al.} 2008{\natexlab{a}}, Nuclear Instruments and Methods in
  Physics Research A, 588, 52

\bibitem[{{Tavani} {et~al.}(2008{\natexlab{b}}){Tavani}, {Barbiellini},
  {Argan}, {Bulgarelli}, {Caraveo}, {Chen}, {Cocco}, {Costa}, {de Paris}, {Del
  Monte}, {di Cocco}, {Donnarumma}, {Feroci}, {Fiorini}, {Froysland},
  {Fuschino}, {Galli}, {Gianotti}, {Giuliani}, {Evangelista}, {Labanti},
  {Lapshov}, {Lazzarotto}, {Lipari}, {Longo}, {Marisaldi}, {Mastropietro},
  {Mauri}, {Mereghetti}, {Morelli}, {Morselli}, {Pacciani}, {Pellizzoni},
  {Perotti}, {Picozza}, {Pontoni}, {Porrovecchio}, {Prest}, {Pucella},
  {Rapisarda}, {Rossi}, {Rubini}, {Soffitta}, {Trifoglio}, {Trois}, {Vallazza},
  {Vercellone}, {Zambra}, {Zanello}, {Giommi}, {Antonelli}, \&
  {Pittori}}]{Tavani2008:Missione}
---. 2008{\natexlab{b}}, \aap, submitted (arXiv:astro-ph/0807.4254v1)

\bibitem[{{Tavecchio} \& {Ghisellini}(2008)}]{Tavecchio2008:blr:cloudy}
{Tavecchio}, F., \& {Ghisellini}, G. 2008, \mnras, 386, 945

\bibitem[{{Tavecchio} {et~al.}(2007){Tavecchio}, {Maraschi}, {Wolter},
  {Cheung}, {Sambruna}, \& {Urry}}]{Tavecchio2007:jet:power}
{Tavecchio}, F., {Maraschi}, L., {Wolter}, A., {Cheung}, C.~C., {Sambruna},
  R.~M., \& {Urry}, C.~M. 2007, \apj, 662, 900

\bibitem[{{Torres} {et~al.}(2001){Torres}, {Romero}, {Combi}, {Benaglia},
  {Andernach}, \& {Punsly}}]{Torres2001:variab:index}
{Torres}, D.~F., {Romero}, G.~E., {Combi}, J.~A., {Benaglia}, P., {Andernach},
  H., \& {Punsly}, B. 2001, \aap, 370, 468

\bibitem[{{Tosti} {et~al.}(2004){Tosti}, {Bagaglia}, {Campeggi}, {Masetti},
  {Monfardini}, {Nicastro}, {Palazzi}, {Pian}, {Sciuto}, {Zerbi}, {Chincarini},
  {Antonelli}, {Conconi}, {Covino}, {Cutispoto}, {Rodono}, \&
  {Molinari}}]{Tosti2004:REM}
{Tosti}, G. {et~al.} 2004, in Presented at the Society of Photo-Optical
  Instrumentation Engineers (SPIE) Conference, Vol. 5492, Ground-based
  Instrumentation for Astronomy. Edited by Alan F. M. Moorwood and Iye
  Masanori. Proceedings of the SPIE, Volume 5492, pp. 689-700 (2004)., ed.
  A.~F.~M. {Moorwood} \& M.~{Iye}, 689--700

\bibitem[{{Ubertini} {et~al.}(2003){Ubertini}, {Lebrun}, {Di Cocco}, {Bazzano},
  {Bird}, {Broenstad}, {Goldwurm}, {La Rosa}, {Labanti}, {Laurent}, {Mirabel},
  {Quadrini}, {Ramsey}, {Reglero}, {Sabau}, {Sacco}, {Staubert}, {Vigroux},
  {Weisskopf}, \& {Zdziarski}}]{Ubertini2003:ibis}
{Ubertini}, P. {et~al.} 2003, \aap, 411, L131

\bibitem[{{Vercellone} {et~al.}(2007){Vercellone}, {Chen}, {Giuliani},
  {Pellizzoni}, {Mereghetti}, {Fornari}, {Caraveo}, {Perotti}, {Fiorini},
  {Bulgarelli}, {Cocco}, {Labanti}, {Marisaldi}, {Fuschino}, {Galli},
  {Gianotti}, {Trifoglio}, {Tavani}, {Pucella}, {D'Ammando}, {Costa}, {Feroci},
  {Donnarumma}, {Pacciani}, {Monte}, {Argan}, {Trois}, {Prest}, {Vallazza},
  {Picozza}, {Lipari}, {Longo}, {Team}, {Giommi}, {Pittori}, {Antonelli},
  {Gasparrini}, {Cutini}, {Fanari}, {Verrecchia}, \&
  {Salotti}}]{Vercellone2007:atel1160}
{Vercellone}, S. {et~al.} 2007, The Astronomer's Telegram, 1160, 1

\bibitem[{{Vercellone} {et~al.}(2008){Vercellone}, {Chen}, {Giuliani},
  {Bulgarelli}, {Donnarumma}, {Lapshov}, {Tavani}, {Argan}, {Barbiellini},
  {Caraveo}, {Cocco}, {Costa}, {D'Ammando}, {Del Monte}, {De Paris}, {Di
  Cocco}, {Evangelista}, {Feroci}, {Fiorini}, {Froysland}, {Fuschino}, {Galli},
  {Gianotti}, {Labanti}, {Lazzarotto}, {Lipari}, {Longo}, {Marisaldi}, {Mauri},
  {Mereghetti}, {Morselli}, {Pacciani}, {Pellizzoni}, {Perotti}, {Picozza},
  {Prest}, {Pucella}, {Rapisarda}, {Soffitta}, {Trifoglio}, {Trois},
  {Vallazza}, {Vittorini}, {Zambra}, {Zanello}, {Pittori}, {Verrecchia},
  {Gasparrini}, {Cutini}, {Giommi}, {Antonelli}, {Colafrancesco}, \&
  {Salotti}}]{Vercellone2008:3C454_ApJ}
---. 2008, \apjl, 676, L13

\bibitem[{{Villata} {et~al.}(2004){Villata}, {Raiteri}, {Aller}, {Aller},
  {Ter{\"a}sranta}, {Koivula}, {Wiren}, {Kurtanidze}, {Nikolashvili},
  {Ibrahimov}, {Papadakis}, {Tosti}, {Hroch}, {Takalo}, {Sillanp{\"a}{\"a}},
  {Hagen-Thorn}, {Larionov}, {Schwartz}, {Basler}, {Brown}, \&
  {Balonek}}]{Villata2004:WEBT:BLLac}
{Villata}, M. {et~al.} 2004, \aap, 424, 497

\bibitem[{{Villata} {et~al.}(2007){Villata}, {Raiteri}, {Aller}, {Bach},
  {Ibrahimov}, {Kovalev}, {Kurtanidze}, {Larionov}, {Lee}, {Leto},
  {L{\"a}hteenm{\"a}ki}, {Nilsson}, {Pursimo}, {Ros}, {Sumitomo}, {Volvach},
  {Aller}, {Arai}, {Buemi}, {Coloma}, {Doroshenko}, {Efimov}, {Fuhrmann},
  {Hagen-Thorn}, {Kamada}, {Katsuura}, {Konstantinova}, {Kopatskaya}, {Kotaka},
  {Kovalev}, {Kurosaki}, {Lanteri}, {Larionova}, {Mingaliev}, {Mizoguchi},
  {Nakamura}, {Nikolashvili}, {Nishiyama}, {Sadakane}, {Sergeev}, {Sigua},
  {Sillanp{\"a}{\"a}}, {Smart}, {Takalo}, {Tanaka}, {Tornikoski}, {Trigilio},
  \& {Umana}}]{vil07}
---. 2007, \aap, 464, L5

\bibitem[{{Villata} {et~al.}(2006){Villata}, {Raiteri}, {Balonek}, {Aller},
  {Jorstad}, {Kurtanidze}, {Nicastro}, {Nilsson}, {Aller}, {Arai}, {Arkharov},
  {Bach}, {Ben{\'{\i}}tez}, {Berdyugin}, {Buemi}, {B{\"o}ttcher}, {Carosati},
  {Casas}, {Caulet}, {Chen}, {Chiang}, {Chou}, {Ciprini}, {Coloma}, {di Rico},
  {D{\'{\i}}az}, {Efimova}, {Forsyth}, {Frasca}, {Fuhrmann}, {Gadway}, {Gupta},
  {Hagen-Thorn}, {Harvey}, {Heidt}, {Hernandez-Toledo}, {Hroch}, {Hu}, {Hudec},
  {Ibrahimov}, {Imada}, {Kamata}, {Kato}, {Katsuura}, {Konstantinova},
  {Kopatskaya}, {Kotaka}, {Kovalev}, {Kovalev}, {Krichbaum}, {Kubota},
  {Kurosaki}, {Lanteri}, {Larionov}, {Larionova}, {Laurikainen}, {Lee}, {Leto},
  {L{\"a}hteenm{\"a}ki}, {L{\'o}pez-Cruz}, {Marilli}, {Marscher}, {McHardy},
  {Mondal}, {Mullan}, {Napoleone}, {Nikolashvili}, {Ohlert}, {Postnikov},
  {Pursimo}, {Ragni}, {Ros}, {Sadakane}, {Sadun}, {Savolainen}, {Sergeeva},
  {Sigua}, {Sillanp{\"a}{\"a}}, {Sixtova}, {Sumitomo}, {Takalo},
  {Ter{\"a}sranta}, {Tornikoski}, {Trigilio}, {Umana}, {Volvach}, {Voss}, \&
  {Wortel}}]{vil06}
---. 2006, \aap, 453, 817

\bibitem[{{Woo} \& {Urry}(2002)}]{Woo2002:BH:masses}
{Woo}, J.-H., \& {Urry}, C.~M. 2002, \apj, 579, 530

\bibitem[{{Zerbi} {et~al.}(2004){Zerbi}, {Chincarini}, {Ghisellini}, {Rodono},
  {Tosti}, {Antonelli}, {Conconi}, {Covino}, {Cutispoto}, {Molinari},
  {Nicastro}, \& {Palazzi}}]{Zerbi2004:REM}
{Zerbi}, F.~M. {et~al.} 2004, in Presented at the Society of Photo-Optical
  Instrumentation Engineers (SPIE) Conference, Vol. 5492, Ground-based
  Instrumentation for Astronomy. Edited by Alan F. M. Moorwood and Iye
  Masanori. Proceedings of the SPIE, Volume 5492, pp. 1590-1601 (2004)., ed.
  A.~F.~M. {Moorwood} \& M.~{Iye}, 1590--1601

\end{thebibliography}

\clearpage

%%%%%%%%%%%%%%%%%%%%%%%%%%%%%%%%%%%%%%%%%%%%%%%%%%%%%%%%%%%%%%%%
%%%%%%%%%%%%%%%%%%%%%%%%%%%%%%%%%%%%%%%%%%%%%%%%%%%%%%%%%%%%%%%%

%
%         TABLE 1 - ISGRI SPECTRAL RESULT
%
\begin{deluxetable}{lrrr}
  \tablecolumns{4}
  \tabletypesize{\normalsize}
  \tablecaption{\igr{}/IBIS spectral fit results. \label{3c454:tab:igr:specfits}} 	
  \tablewidth{0pt}
  \tablehead{
    \colhead{Rev.}  & \colhead{$\Gamma$}  & \colhead{$\chi^2_{\rm red}$ (d.o.f.)} & \colhead{Flux}\tablenotemark{a}}
  \startdata
  623       & $1.78_{-0.30}^{+0.33}$  & 1.21 (11) & $1.52$ \\
  624       & $1.71_{-0.36}^{+0.41}$  & 0.54 (11) & $1.42$ \\
  623$+$624 & $1.75_{-0.23}^{+0.25}$  & 0.89 (11) & $1.49$ \\
  \enddata
  \tablenotetext{(a)}{Flux in the 20--200\,keV band in units of 
    $10^{-3}$\phcmsec obtained from the spectral fits.}
\end{deluxetable} 
%

%
%        TABLE 2 - XRT SPECTRAL RESULT
%
\begin{deluxetable}{lllccc} 	
  \tablecolumns{6} 
  \tabletypesize{\normalsize}
  \tablecaption{Swift/XRT spectral fit results.\label{3c454:tab:xrt:specfits}}
  \tablewidth{0pt}
  \tablehead{
    \colhead{Obs.\tablenotemark{a}} & \colhead{$N_{\rm H}$}           & \colhead{$\Gamma$}  & \colhead{$\chi^2_{\rm red}/$(d.o.f.)} & \colhead{$F_{\rm 0.3-10}$ (obs.)}  & \colhead{$F_{\rm 2-10}$ (obs.)}\\
    \colhead{} & \colhead{(10$^{22}$ cm$^{2}$)}  & \colhead{}          &  \colhead{}                 &  \colhead{\phcmsec}         & \colhead{\phcmsec}}
  \startdata
  \cutinhead{Model A: single power law with Galactic$+$intrinsic absorption}\\
  001       & $0.0724$+$(0.06_{-0.06}^{+0.06})$ & $1.56_{-0.07}^{+0.07}$  & $1.173/(108)$ & $1.45 \times 10^{-2}$ & $0.52 \times 10^{-2}$\\
  003$+$004 & $0.0724$+$(0.20_{-0.10}^{+0.11})$ & $1.68_{-0.10}^{+0.11}$  & $0.933/(53)$  & $1.41 \times 10^{-2}$ & $0.51 \times 10^{-2}$\\
  005       & $0.0724$+$(0.23_{-0.12}^{+0.14})$ & $1.60_{-0.12}^{+0.12}$  & $0.900/(46)$  & $1.43 \times 10^{-2}$ & $0.56 \times 10^{-2}$\\
  006       & $0.0724$+$(0.25_{-0.78}^{+0.88})$ & $1.68_{-0.09}^{+0.09}$  & $1.089/(80)$  & $1.28 \times 10^{-2}$ & $0.48 \times 10^{-2}$\\
  \cutinhead{Model B: single power law with free absorption}\\
  001	    & $0.10_{-0.02}^{+0.02}$  &  $1.56_{-0.07}^{+0.07}$  &  $1.100/(131)$  &  $1.39\times 10^{-2}$ &  $0.51\times 10^{-2}$ \\
  003$+$004 & $0.15_{-0.03}^{+0.04}$  &  $1.73_{-0.12}^{+0.13}$  &  $0.921/(53)$   &  $1.40\times 10^{-2}$ &  $0.50\times 10^{-2}$ \\
  005       & $0.15_{-0.03}^{+0.04}$  &  $1.66_{-0.11}^{+0.12}$  &  $1.070/(58)$   &  $1.35\times 10^{-2}$ &  $0.52\times 10^{-2}$ \\ 
  006       & $0.16_{-0.02}^{+0.03}$  &  $1.72_{-0.09}^{+0.09}$  &  $1.124/(94)$   &  $1.23\times 10^{-2}$ &  $0.46\times 10^{-2}$ \\ 
  \enddata
  \tablenotetext{(a)}{Last three digits of observation number 00031018}
\end{deluxetable} 
%

%
%         TABLE 3 - SED SPECTRAL RESULT
%
\begin{deluxetable}{lrrr} 	
  \tablecolumns{4}
  \tabletypesize{\normalsize}
  \tablecaption{Input parameters for the model of P1 and P2 SEDs.
    See text for details.  \label{3c454:tab:sed:param}} 	
  \tablewidth{0pt}
  \tablehead{
    \colhead{Parameter} & \colhead{SED P1} & \colhead{SED P2} & \colhead{Units}}
  \startdata
  $\alpha_{\rm l}$             & 2.1   & 2.2   & \\
  $\alpha_{\rm h}$             & 4.5   & 5.0   & \\
  $\gamma_{\rm min}$           & 10    & 10    & \\
  $\gamma_{\rm b}$             & 500   & 500   & \\
  $K$                          & 14    & 12    & cm$^{-3}$ \\
  $R$                          & 35    & 35    & 10$^{15}$\,cm\\
  $B$                          & 10    & 8     & G\\
  $\delta$                     & 14.64 & 14.64 & \\
  $L_{\rm d}$                  & 5     & 5     & 10$^{46}$\,erg\,s$^{-1}$\\
  $r$                          & 0.05  & 0.05  & pc\\
  $\Theta_{0}$                 & 2.6   & 2.6   & degrees\\
  $\Gamma$                     & 8.4   & 8.4   & \\
  \enddata
\end{deluxetable} 

\clearpage%%%%%%%%%%%%%%%%%%%%%%%%%%%%%%%%%%%%

%
%------------------ FIG 1 : CTS MAP -------------------------
%-------------------------------------------------------------
%

%----------------- FIG 2 : LONG TERM LC ---------------------
%-------------------------------------------------------------
%

%
%----------------- FIG 3 : DAY.BY.DAY LC --------------------
%-------------------------------------------------------------
%

%
%----------------- FIG 4 : GRID SPECTRUM --------------------
%-------------------------------------------------------------
%

%----------------- FIG 5 : ISGRI LC --------------------------
%-------------------------------------------------------------
%

%
%------------------ FIG 6 : ISGRI SPECTRUM -------------------
%-------------------------------------------------------------
%

%
%------------------- FIG 7 : XRT SPECTRA ---------------------
%-------------------------------------------------------------
%

%
%------------------ FIG 8 : XRT FLUX VS GAMMA ----------------
%-------------------------------------------------------------
%

%
%----------------- FIG 9 : BAT LC ----------------------------
%-------------------------------------------------------------
%

%
%---------------- FIG 10 : MULTILAMBDA LC --------------------
%-------------------------------------------------------------
%

%
%--------------- FIG 11 : DCF --------------------------------
%-------------------------------------------------------------
%

%
%--------------- FIG 12 : SED P1 -----------------------------
%-------------------------------------------------------------
%

%
%--------------- FIG 13 : SED P2 -----------------------------
%-------------------------------------------------------------
%

%
%--------------- FIG 14 : MULTI SPECTRA ----------------------
%-------------------------------------------------------------
%

%

%%%%%%%%%%%%%%%%%%%%%%%%%%%%%%%%%%%%%%%%%%%%%%%%%%%%%%%%%%%%%%%%%%%%
%%%%%%%%%%%%%%%%%%%%%%%%%%%%%%%%%%%%%%%%%%%%%%%%%%%%%%%%%%%%%%%%%%%%
%%%%%%%%%%%%%%%%%%%%%%%%%%%%%%%%%%%%%%%%%%%%%%%%%%%%%%%%%%%%%%%%%%%%
\end{document}